\documentclass[reprint,groupedaddress,amsmath,amssymb,aps,prl,floatfix,longbibliography]{revtex4-1}
\usepackage{graphicx}
\graphicspath{{figures/}}
\usepackage{amsmath}
\usepackage{amssymb}
\usepackage{multirow} 
\usepackage{physics} 
\usepackage{bbold}
\usepackage{xcolor}

\usepackage[section]{placeins}

\graphicspath{{figures/}}
\newcommand{\pvec}[1]{\vec{#1}\mkern2mu\vphantom{#1}}

\AtBeginDocument{\let\oldcontentsline\contentsline}
\newcommand{\notoccontentsline}[4]{\oldcontentsline{}{}{}{}}
\newcommand{\droptocpage}{\addtocontents{toc}{\let\protect\contentsline\protect\notoccontentsline}}
\newcommand{\incltocpage}{\addtocontents{toc}{\let\protect\contentsline\protect\oldcontentsline}}

\begin{document}
\title{A tweezer clock with half-minute atomic coherence at optical frequencies and high relative stability}
\author{Aaron W. Young}
\author{William J. Eckner}
\author{William R. Milner}
\author{Dhruv Kedar}
\author{Matthew A. Norcia}
\author{Eric Oelker}
\author{Nathan Schine}
\author{Jun Ye}
\author{Adam M. Kaufman}
\email[E-mail: ]{adam.kaufman@colorado.edu}
\address{JILA, University of Colorado and National Institute of Standards and Technology, and Department of Physics, University of Colorado, Boulder, Colorado 80309, USA}

\begin{abstract} 

The preparation of large, low-entropy, highly coherent ensembles of identical quantum systems is foundational for many studies in quantum metrology, simulation, and information. Here, we realize these features by leveraging the favorable properties of tweezer-trapped alkaline-earth atoms while introducing a new, hybrid approach to tailoring optical potentials that balances scalability, high-fidelity state preparation, site-resolved readout, and preservation of atomic coherence. With this approach, we achieve trapping and optical clock excited-state lifetimes exceeding $ 40 $ seconds in ensembles of approximately $ 150 $ atoms. This leads to half-minute-scale atomic coherence on an optical clock transition, corresponding to quality factors well in excess of $10^{16}$. These coherence times and atom numbers reduce the effect of quantum projection noise to a level that is on par with leading atomic systems, yielding a relative fractional frequency stability of $5.2(3)\times10^{-17}~(\tau/s)^{-1/2}$ for synchronous clock comparisons between sub-ensembles within the tweezer array. When further combined with the microscopic control and readout available in this system, these results pave the way towards long-lived engineered entanglement on an optical clock transition in tailored atom arrays.

\end{abstract}

\date{\today}

\maketitle

A key requirement in quantum metrology, simulation, and information is the control and preservation of coherence in large ensembles of effective quantum two level systems, or qubits~\cite{preskill_quantum_2012, georgescu_quantum_2014, ludlow_optical_2015}. One way to realize these features is with neutral atoms~\cite{saffman_quantum_2010, browaeys_many-body_2020}, which benefit from being inherently identical, and having weak and short-range interactions in their ground states. This, combined with the precise motional and configurational control provided by tailored optical potentials, enables assembly of large ensembles of atomic qubits~\cite{barredo_atom-by-atom_2016, endres_atom-by-atom_2016, kumar_sorting_2018, brown_gray-molasses_2019} without the need for careful calibration of individual qubits or additional shielding from uncontrolled interactions with the environment. As a result, groundbreaking work has been done in such systems using alkali atoms, including the realization of controllable interactions and gates~\cite{levine_parallel_2019, graham_rydberg-mediated_2019}, preparation of useful quantum resources~\cite{omran_generation_2019}, and simulation of various spin models of interest~\cite{bernien_probing_2017, leseleuc_observation_2019}. These techniques have recently been extended to alkaline-earth (or alkaline-earth-like) atoms~\cite{cooper_alkaline-earth_2018, norcia_microscopic_2018, saskin_narrow-line_2019}, which further provide access to extremely long-lived nuclear and electronic excited states, and new schemes for Rydberg spectroscopy~\cite{wilson_trapped_2019, madjarov_high-fidelity_2020-1}.

These advancements have enabled the development of tweezer-array optical clocks~\cite{madjarov_atomic-array_2019, norcia_seconds-scale_2019-1}. These clocks leverage the flexible potentials provided by optical tweezer arrays to rapidly prepare and interrogate ensembles of many non-interacting atoms, and, consequently, balance the pristine isolation and high duty cycles available in single ion-based optical clocks~\cite{chou_frequency_2010, brewer_$27mathrm+$_2019} with the large ensembles and resultant low quantum projection noise (QPN) available in optical lattice clocks~\cite{ludlow_optical_2015, ushijima_cryogenic_2015, campbell_fermi-degenerate_2017-1, oelker_demonstration_2019}. The most stable tweezer clock demonstrated to date used a one-dimensional (1D) array containing 5 atoms, and consequently was limited by QPN to a stability of $4.7\times10^{-16}~(\tau/s)^{-1/2}$~\cite{norcia_seconds-scale_2019-1}, about an order of magnitude worse than the record values of $3.1\times10^{-17}~(\tau/s)^{-1/2}$ reported for synchronous comparisons in a 3D lattice clock~\cite{campbell_fermi-degenerate_2017-1}, and $4.8\times10^{-17}~(\tau/s)^{-1/2}$ for a comparison between two clocks~\cite{oelker_demonstration_2019}. Extending tweezer-array clocks to large 2D arrays would help to close this gap by increasing atom number while maintaining the high duty cycles achievable in tweezer-based systems~\cite{norcia_seconds-scale_2019-1}.

This tweezer-clock architecture also benefits from microscopic single-particle control through 100-nanometer-precision positioning of individual atoms, which can be leveraged to protect quantum coherence. The importance of such capabilities to optical lattice clocks was recently illuminated by Hutson et al.~\cite{hutson_engineering_2019}: in 3D lattice clocks, record atomic coherence times and stabilities are set by a balance between suppressing atomic tunneling and lattice-induced spontaneous Raman scattering~\cite{campbell_fermi-degenerate_2017-1, oelker_demonstration_2019}. In a fixed-wavelength lattice these two effects are coupled through the trap depth, with reduced tunneling in deeper traps, and reduced scattering in shallower traps, leading to an optimum. In a tweezer array, the tunneling and spontaneous Raman scattering can be controlled independently via the atomic spacing and the tweezer depth, allowing for simultaneous suppression of both effects and potentially extremely long coherence times.

\begin{figure*}[!t]
	\includegraphics[width=\linewidth]{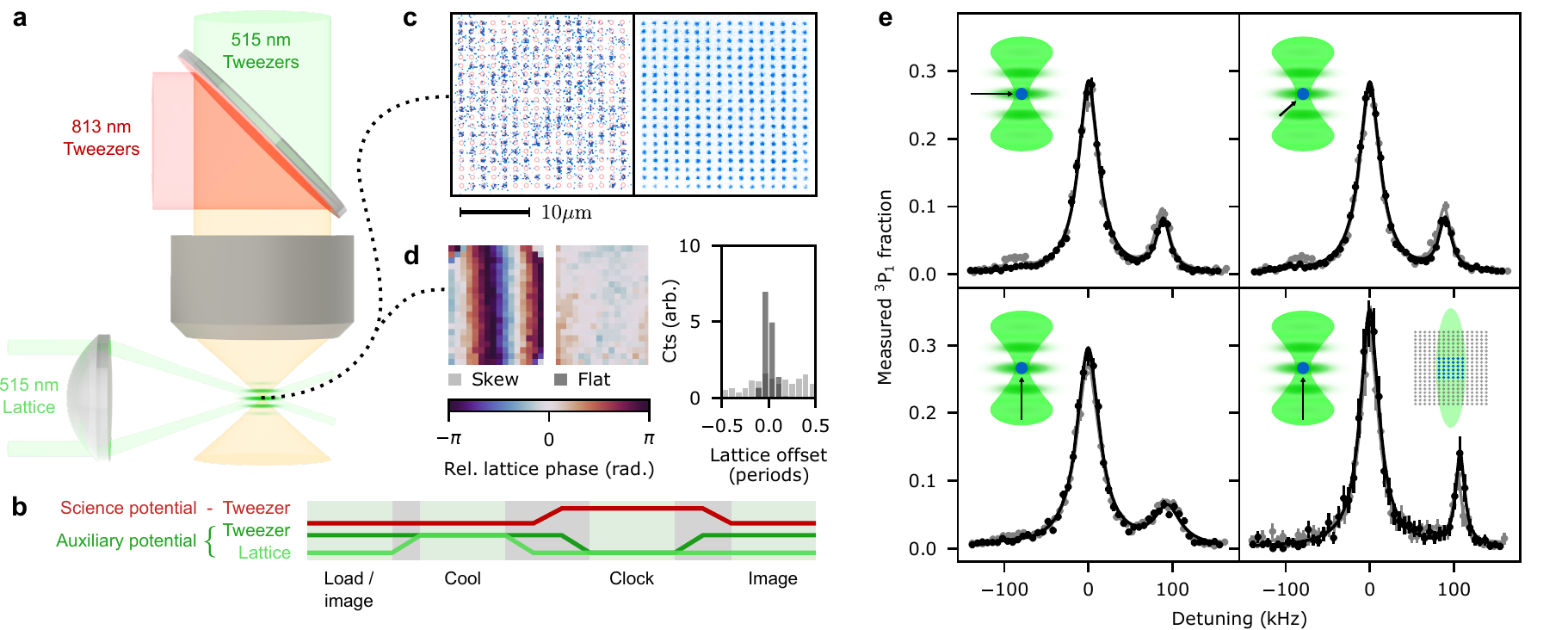}
	\caption{\textbf{3D ground-state cooled strontium atoms in a 320-site clock-magic wavelength tweezer array.} a) In order to generate large numbers of traps that are compatible with ground-state cooling and narrow-line spectroscopy, we combine a shallow clock-magic ``science'' potential at 813~nm with a tightly confining ``auxiliary'' potential at 515~nm. The auxiliary potential includes a tweezer array and a crossed-beam optical lattice to provide tight confinement along all spatial axes. b) In a typical experimental sequence these potentials cooperate to prepare and readout 3D ground-state cooled atoms in traps that are compatible with narrow-line clock spectroscopy. c) Representative single shot (left) and averaged (right) images of atoms demonstrate site-resolved readout of the $16\times20$ array of tweezers used in this work, with a spacing of 1.2~$\mathrm{\mu m}$ (1.5~$\mathrm{\mu m}$) in the vertical (horizontal) direction. The red circles in the single-shot image denote the tweezer positions to guide the eye. d) Measurements of the spatial phase of the standing-wave lattice at each tweezer~\cite{Note1} with an intentional tilt (left) and properly aligned (center) show that it is possible flatten the lattice relative to the entire tweezer array to within $\mathrm{1/10^{th}}$ of a lattice period (histogram, right). This allows for high-fidelity sideband cooling in all axes. e) Performing sideband spectroscopy before (black points) and after (grey points) adiabatically transferring the atoms to and back from the science potential, we measure an average phonon occupation of $\Bar{n} = 0.07^{+0.14}_{-0.07} $, $ 0.06^{+0.08}_{-0.06} $, and $ 0.07\pm0.06 $ ($\Bar{n} = 0.25\pm 0.12 $, $ 0.31\pm 0.13 $, and $ 0.27\pm 0.10 $) before (after) the handoff in the axial, first, and second radial directions respectively. Cartoons in the top left of each frame indicate the orientation of the probe beam relative to the traps, showing probes in two orthogonal radial directions (top) and in the axial direction (bottom). The bottom-right spectra show that, in a reduced $6\times6$ region at the center of the array (denoted by the bottom-right cartoon), the axial cooling performance is vastly improved, with an average phonon occupation of $\Bar{n} = 0.00^{+0.06}_{-0.00} $ ($\Bar{n} = 0.06^{+0.10}_{-0.06} $) before (after) the handoff. This is due to the comparable extent of the lattice beams to the tweezer array (light-green contour in bottom-right cartoon shows region over which the lattice intensity stays within 90\% of its maximal value).}
	\label{fig:setup}
\end{figure*}

Tweezer clocks are also attractive for quantum metrology and simulation based on the use of Rydberg-mediated interactions in programmable alkaline-earth atom arrays~\cite{gil_spin_2014, kaubruegger_variational_2019}. The large 2D arrays and tight spacings used in this work are key for future studies involving limited-range Rydberg interactions, providing access to larger samples with higher connectivity, stronger interactions, and correspondingly greater entanglement. Furthermore, while many-body entanglement scales exponentially poorly with single-particle decoherence, the coherence times reported below establish the prospect of a metrologically useful entangled optical clock operating with tens of atoms and seconds-long interrogation times. Our use of $\mathrm{^{88}Sr}$, whose clock linewidth is tunable with a magnetic field, also establishes longer-term directions for quantum metrology that are not fundamentally limited by spontaneous emission~\cite{kessler_heisenberg-limited_2014}.

Through a series of advances in this work, we show sub-hertz control of an optical clock transition in a tweezer array of 320 traps containing a total of on average $\sim150$ atoms (see Fig.~\ref{fig:setup}a-d). We demonstrate the ability to load ground-state cooled atoms into shallow clock-magic tweezers, achieving excited-state lifetimes of up to 46(5) seconds and homogeneity on the scale of tens of millihertz. As a consequence, we measure a coherence time of 19.5(8)~s for synchronous frequency comparisons involving the entire array, and observe evidence of atomic coherence out to 48(8) seconds for select atoms in the array, corresponding to an atomic quality factor ($Q$) of $6.5(1.1)\times10^{16}$. These characteristics reduce the effects of QPN in the tweezer clock platform to a level that is on par with the state of the art~\cite{campbell_fermi-degenerate_2017-1, oelker_demonstration_2019}, yielding a relative fractional frequency stability of 5.2(3)$\times10^{-17}~(\tau/s)^{-1/2}$ for synchronous self-comparisons.

A central challenge for using tweezer-array systems in quantum science is maintaining control while scaling to larger atom numbers. Fundamentally, given finite optical power, an increase in sample size comes at the expense of trap depth and atomic confinement, with implications for detection fidelity, cooling performance, qubit coherence, and atomic loading. Such trade-offs impact the viability of the platform for quantum information, quantum simulation, and metrology. In this work, we address this challenge in the context of the tweezer clock, but our approach has general relevance to these other endeavors in quantum science. 

Our solution is to use several optical potentials optimized for different stages of the experiment, and to realize state-preserving, low-loss transfer between these different potentials~\cite{liu_molecular_2019}. We use an ``auxiliary'' potential for initial state preparation and readout and a ``science'' potential for clock interrogation (see Fig.~\ref{fig:setup}a,~b). The auxiliary potential includes a 2D tweezer array and a crossed-beam optical lattice, which provides additional confinement along the weakly confined ``axial'' axis of the tweezers. Because the required confinement is the same in all axes for 3D ground-state cooling, this axial lattice greatly reduces the power requirements on the auxiliary tweezers. In our apparatus, with a numerical aperture of $\mathrm{NA}\simeq0.68$, this corresponds to a $\sim30$-fold reduction in required optical power per tweezer. As a result, at modest optical power, we can create near-spherical traps with roughly 90~kHz trap frequencies in all axes. Including various losses in our system, and using $\sim4$~W of total optical power, we create 320 such traps in a $16\times20$ array, with 1.5 and 1.2~$\mathrm{\mu m}$ spacings along the two axes (Fig.~\ref{fig:setup}c,~e).

The science potential is a 2D tweezer array operating at 813~nm, a magic wavelength for the clock transition~\cite{norcia_seconds-scale_2019-1}, whereas the auxiliary potential operates at 515~nm, where a magic trapping condition can be achieved for the $\mathrm{^1S_0} \Leftrightarrow \mathrm{^3P_1}$ cooling transition at 689~nm via tuning of a magnetic field~\cite{norcia_microscopic_2018}. The power requirements at 813~nm are more demanding compared to 515~nm, due to the roughly $3\times$ lower polarizability, larger diffraction-limited spot size, and reduction in available laser power at this wavelength. However, critically, because the science potential is only used for the clock-interrogation stage where shallow traps are preferable, these power constraints do not impose a limitation on atom number or state preparation.   

\begin{figure}[!t]
    \includegraphics[width=\linewidth]{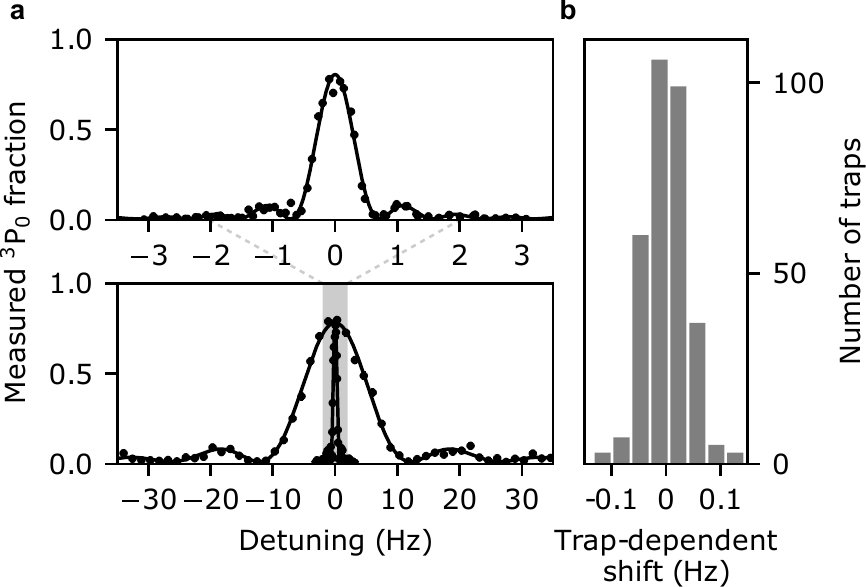}
    \caption{\textbf{Sub-Hertz clock spectroscopy in 320 tweezers.} a) Array-averaged Rabi spectroscopy of the $\mathrm{^1S_0} \Leftrightarrow \mathrm{^3P_0}$ clock transition provides Fourier-limited linewidths of 10.1(2)~Hz and 0.62(1)~Hz (full width at half maximum), in good agreement with sinc lineshapes generated from the known probe durations used in each case (solid lines). Callout (top) shows the Fourier-limited 0.6~Hz feature in detail, with no reduction in transfer fraction compared to the 10~Hz case. Error bars are smaller than the point size. b) We investigate the presence of inhomogeneous, trap-dependent shifts of the clock transition by performing site-resolved spectroscopy. The fitted centers of these spectra have a standard deviation of 0.039(2)~Hz.}
    \label{fig:rabi}
\end{figure}

\begin{figure*}[!t]
    \includegraphics[width=\linewidth]{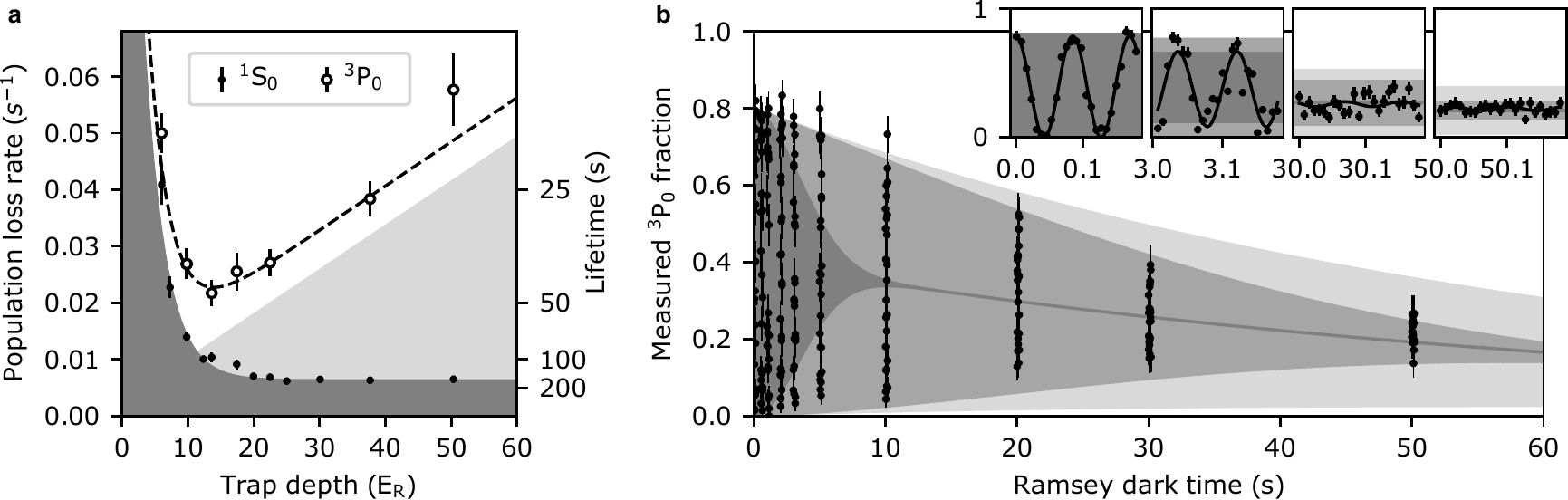}
    \caption{\textbf{Minute-scale atomic lifetime and coherence.} a) To determine limits on atomic coherence, we measure the lifetime of both the ground ($\mathrm{^1S_0}$, black points) and clock ($\mathrm{^3P_0}$, black circles) states. For ground-state atoms the lifetime saturates to 160(10) seconds in deep traps, with additional technical sources of atom loss contributing in shallower traps (exponential fit to $\mathrm{^1S_0}$ data, dark grey). For clock-state atoms an optimal trap depth arises from a competition between this atom loss, which prefers deep traps, and depumping via spontaneous Raman scattering of the trap light (theory prediction with no free parameters, light grey)~\cite{dorscher_lattice-induced_2018, hutson_engineering_2019, Note1}, which prefers shallow traps. The combination of these loss mechanisms (dashed line) is in good agreement with the measured clock-state lifetimes, including the optimum of 46(5) seconds at 14$\mathrm{E_R}$. b) For clock operation, we perform Ramsey spectroscopy in 15$\mathrm{E_R}$ deep tweezers (black points), near this optimal depth. Given the measured clock-state lifetime, we would expect the contrast to decay with an exponential time constant of 55(8)~s (light-grey region). However, we expect tweezer-dependent light shifts to result in Gaussian decay with a time constant of 33(1)~s at 15$\mathrm{E_R}$~\cite{Note1, norcia_seconds-scale_2019-1}. The combination of these two effects is denoted by the medium-grey region. Note that each data point corresponds to a single shot of the experiment. As a result, despite the fact that the atom-laser coherence decays with a Gaussian time constant of 3.6(2)~s (dark-grey region)~\cite{Note1}, the variance of the Ramsey signal decays on a timescale set by atomic coherence. This is clarified by the insets, which share units with the main axes, and show detailed views of Ramsey evolution at a few different times. Here, it is possible to see the initial loss of phase coherence with the laser followed, at later times, by total loss of coherence (and thus variance in this signal) in the system.}
    \label{fig:lifetime}
\end{figure*}

To perform clock spectroscopy the 813~nm tweezers are adiabatically ramped on, and the 515~nm tweezers and lattice ramped off. After exciting some atoms to the clock state, the atoms are adiabatically transferred back into the 515~nm tweezers for final readout. With optimal alignment, this whole ``handoff'' procedure can be performed with 0.0(3)\% additional atom loss~\footnote{See Supplemental Material at [URL will be inserted by publisher]}. However, this platform is currently limited by loss during imaging in the 515~nm tweezers, contributing to a background of 5\% loss per image pair in our apparatus~\cite{cooper_alkaline-earth_2018, norcia_microscopic_2018}.

To confirm that the atoms remain cold during the handoff we perform sideband thermometry via the procedure described in our previous work~\cite{norcia_microscopic_2018}. This is done in the auxiliary potential (including the lattice) immediately after sideband cooling, and after adiabatically passing the atoms to the science potential, holding for 25~ms, and passing them back~\cite{Note1}. As shown in Fig.~\ref{fig:setup}e, before the handoff we observe an average phonon occupation of $\Bar{n} = 0.07^{+0.14}_{-0.07} $, $ 0.06^{+0.08}_{-0.06} $, and $ 0.07\pm0.06 $ in the axial, first, and second radial directions respectively. After the handoff we observe an average phonon occupation of $\Bar{n} = 0.25\pm 0.12 $, $ 0.31\pm 0.13 $, and $ 0.27\pm 0.10 $ (again in the axial, first, and second radial directions). Since we expect that heating occurs during both steps of the handoff, the mean of these two measurements serves as an estimate of the temperature of the atoms in the science potential.

While the tweezers, and thus the radial trap frequencies, can be balanced across the entire array, there is substantial inhomogeneous broadening of the axial trap frequencies. This is due to the relatively small $\mathrm{25~\mu m}$ waists of the lattice beams, which are comparable to the extent of the tweezer array (Fig.~\ref{fig:setup}e). In a smaller $ 6\times6 $ region at the center of the array the axial cooling and handoff performance is vastly improved, with an average phonon occupation of $\Bar{n} = 0.00^{+0.06}_{-0.00} $ ($\Bar{n} = 0.06^{+0.10}_{-0.06} $) before (after) the handoff. Due to the modest power requirements of the lattice, the lattice waist could easily be increased in the future without sacrificing axial trap frequency, suggesting that this enhanced performance could be achieved across the entire array.

After loading ground-state cooled atoms into the science potential, we can interrogate the clock transition. As in our previous work~\cite{norcia_seconds-scale_2019-1}, we apply a magnetic field of 22~G to mix the $\mathrm{^3P_1}$ state into the $\mathrm{^3P_0}$ state which opens the doubly forbidden $\mathrm{^1S_0} \Leftrightarrow \mathrm{^3P_0}$ transition at 698~nm~\cite{taichenachev_magnetic_2006}. By applying laser light that is referenced to an ultra-stable crystalline cavity~\cite{oelker_demonstration_2019} and resonant with this transition we excite atoms to the $\mathrm{^3P_0}$ state. To detect the population excited to $\mathrm{^3P_0}$ we apply a ``blow-away'' pulse of 461~nm light resonant with the $\mathrm{^1S_0} \Leftrightarrow \mathrm{^1P_1}$ transition to remove atoms that were in the ground state. The remaining atoms are adiabatically transferred back into the 515~nm tweezers, repumped to the ground state, and imaged (see methods). With this protocol we observe no reduction in $\mathrm{^3P_0}$ transfer fraction when using Rabi frequencies between $2\pi\times7$~Hz and $2\pi\times0.4$~Hz averaged across all $\sim150$ atoms in the array (Fig.~\ref{fig:rabi}a).

All Rabi spectroscopy is performed in 25$\mathrm{E_R}$ deep tweezers (where $\mathrm{E_R}$ is the single photon recoil energy), corresponding to 58~$\mathrm{\mu W}$ of optical power per tweezer as measured at the atoms. These shallow traps are the primary limit on transfer fraction for all Rabi frequencies used in this work. Specifically, these depths result in a relatively high Lamb-Dicke parameter of $\eta = 0.83$, and thus increased sensitivity to residual motional excitation~\cite{Note1}. However, the benefit of using such shallow traps is that clock frequency shifts arising from spatial variation of the tweezer wavelengths should be bounded to below 50~mHz across the entire array, resulting in reduced dephasing~\cite{norcia_seconds-scale_2019-1}. To confirm this, we fit the clock transition frequency at each tweezer, and measure a standard deviation in trap-dependent clock frequencies of 39(2)~mHz (Fig.~\ref{fig:rabi}b).

The ability to operate at these shallow depths is, in part, due to the flexibility afforded by independently optimizing two separate tweezer systems for shallow and deep operation. This extra freedom makes it possible to minimize various technical sources of heating and atom loss in shallow tweezers~\cite{Note1}. As a result, we observe lifetimes of 160(10)~s down to 25$\mathrm{E_R}$ (Fig.~\ref{fig:lifetime}a) --- a quarter of the shallowest depths reported in previous works~\cite{norcia_seconds-scale_2019-1} --- likely limited by our vacuum. 

To maximize clock-state coherence, it is desirable to go to even lower depths to reduce the effect of decoherence via Raman-transitions driven by the trap photons~\cite{dorscher_lattice-induced_2018, hutson_engineering_2019}. Unlike in lattice clocks, where the effects of tunneling can become limiting at depths below $\sim30\mathrm{E_R}$ along a single axis ($\sim100\mathrm{E_R}$ in a 3D lattice)~\cite{hutson_engineering_2019}, we observe no evidence of tunneling or thermal hopping in tweezers as shallow as 6$\mathrm{E_R}$~\cite{Note1}. Importantly, at this depth we calculate the tunneling rate to be $ \sim1$~Hz, suggesting that disorder also plays a key role in pinning the atoms. While this is encouraging, at these depths technical sources of atom loss~\cite{Note1} begin to limit our trap lifetime to far below 160(10)~s. A competition between these losses and Raman scattering leads to an optimal trap depth of $\sim14\mathrm{E_R}$, where we measure an excited clock-state lifetime of 46(5)~s (Fig.~\ref{fig:lifetime}a). This lifetime is in good agreement with the predicted value of 44(6)~s based on the measured ground-state trap lifetime of 96(8)~s, and the expected contributions from trap induced Raman scattering and black-body radiation~\cite{dorscher_lattice-induced_2018, hutson_engineering_2019}.

Our measured lifetimes suggest that, at 15$\mathrm{E_R}$, Ramsey contrast should decay exponentially with a time constant of 55(8)~s. In practice, this decay is exacerbated by tweezer-induced frequency shifts~\cite{madjarov_atomic-array_2019, norcia_seconds-scale_2019-1}, which we expect to result in Gaussian decay with a time constant of 33(1)~s~\cite{Note1}. In our measurements, the signal at each Ramsey time is a single-shot measurement such that even though atom-laser coherence decays over $\sim3$~s~\cite{Note1}, we can observe a signal whose variance remains high on much longer timescales (Fig.~\ref{fig:lifetime}b). At short times, the frequency of the Ramsey fringes is set by the differential light-shift imposed by the probe beam on the $\mathrm{^1S_0}$ and $\mathrm{^3P_0}$ states. At longer times, the loss of atom-laser coherence manifests as a randomised phase of the second $\pi/2$ pulse in the Ramsey sequence. This obscures Ramsey oscillations but preserves the probability of large excursions due to the persistence of atomic coherence, where atomic coherence is defined as the magnitude of the off-diagonal elements in the average single-particle density matrix. The Ramsey contrast inferred from this measurement decays with a $1/e$ time of 19.5(8)~s (Fig.~\ref{fig:lifetime}b), slightly faster than the prediction based on the measured lifetime and dephasing. This corresponds to an effective quality factor of $Q=1.9(1)\times10^{16}$ which is limited by inhomogeneous broadening.

\begin{figure}[!t]
    \includegraphics[width=\linewidth]{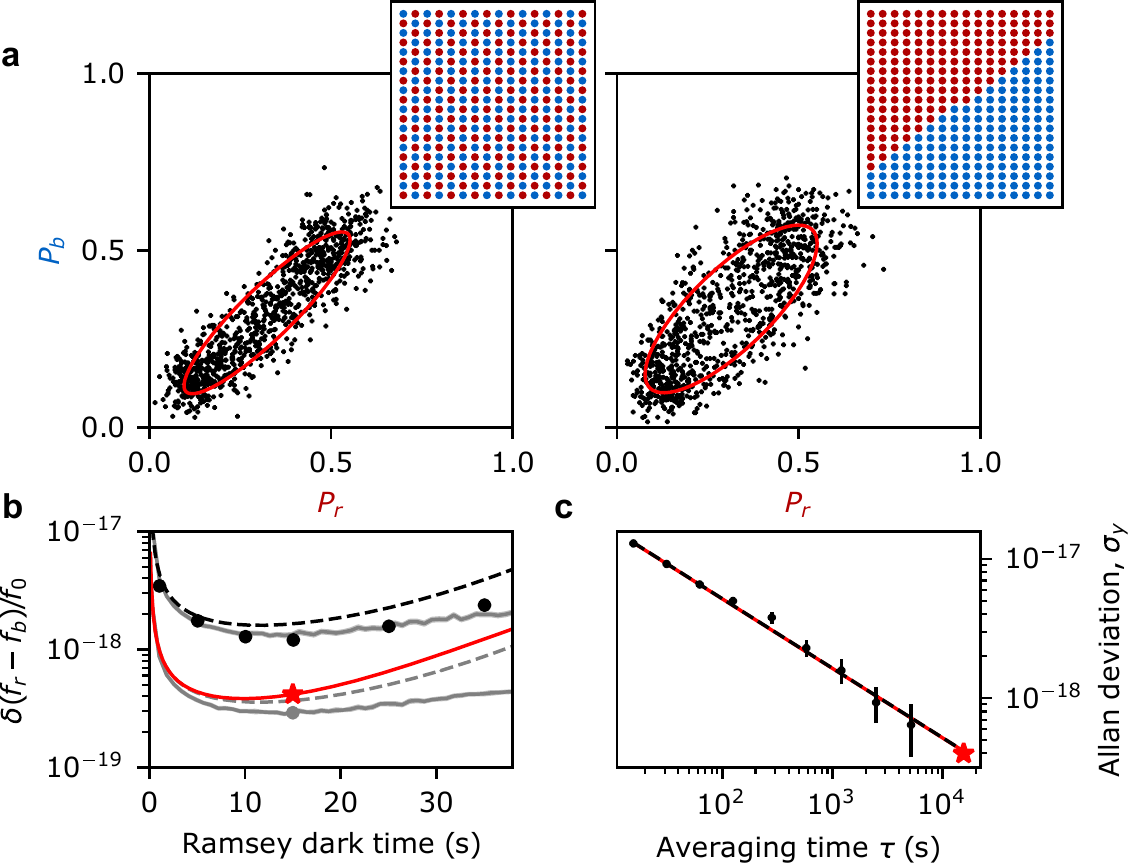}
    \caption{\textbf{Resolving millihertz shifts of an optical transition.} a) We perform a synchronous clock comparison by partitioning the array into two sub-ensembles (insets, red and blue), and creating a parametric plot of the $\mathrm{^3P_0}$ excited-state fraction in the blue ensemble ($P_b$) vs in the red ensemble ($P_r$) (in this case at a 15~s interrogation time in 15$\mathrm{E_R}$ deep tweezers). In the checkerboard (left) partitioning there is no mean frequency shift between the two sub-ensembles, whereas in the diagonal (right) case we expect a 7.0(1.3)~mHz shift~\cite{Note1}. The relative frequency between the sub-ensembles can be extracted via ellipse fitting (red lines), which in the diagonal case yields 7.15(18)~mHz. Note that such fits are biased near zero phase shift, as is evident in the fit to the checkerboard ensemble, which returns an artificially large phase shift. b) To identify an optimal dark time, we compute the fractional frequency uncertainty between the sub-ensembles as a function of Ramsey dark time~\cite{Note1}. The black points (grey point) correspond(s) to 13 minutes (4.3 hours) of averaging, and are extracted from the checkerboard partitioning. Note that these values are not representative of a true stability due to biasing. This is made clear by the dashed curves, which correspond to expected QPN, and the solid grey curves, which include an additional correction factor calculated via Monte-Carlo simulations to account for the biased fits~\cite{Note1} (shaded regions denote 1-sigma confidence interval). At 15~s interrogation times the diagonally separated sub-ensembles have a sufficient phase shift to remove the bias in the fits. This condition (red star) shows the fractional frequency uncertainty of the full 4.3 hour-long measurement, with a value of $4.2\times10^{-19}$. This is in good agreement with the expected QPN limit with no bias correction (red curve). c) We can further compute an Allan deviation associated with this measurement (black points), which averages down with a slope of 5.2(3)$\times10^{-17}~(\tau/s)^{-1/2}$ (black dashed line). This is in good agreement with the expected value of 5.2$\times10^{-17}~(\tau/s)^{-1/2}$ from QPN (red line). Red star is duplicated here as a point of comparison (note that this point is not strictly an Allan deviation, and is extracted via jackknifing~\cite{Note1}).}
    \label{fig:atomatom}
\end{figure}

Even in the absence of atom-laser coherence, we can perform a synchronous clock comparison that takes advantage of this long-lived atomic coherence by comparing the relative phase between two sub-ensembles in the tweezer array~\cite{takamoto_frequency_2011, marti_imaging_2018}. Because readout occurs in a site-resolved manner, the partitioning of these ensembles can be chosen arbitrarily. Specifically, we choose a ``checkerboard'' partitioning that yields no net tweezer-induced frequency shift between the two sub-ensembles, and a ``diagonal'' partitioning that yields a near-maximal frequency shift (Fig.~\ref{fig:atomatom}a insets). As described above, at probe times that exceed the atom-laser coherence time the Ramsey phase is randomized. As a result, parametric plots of the excitation fraction in the two sub-ensembles result in points that fall along an ellipse, where the size of the ellipse is related to the average atomic coherence, and the opening angle of the ellipse is related to the net phase (and thus frequency) shift between sub-ensembles (Fig.~\ref{fig:atomatom}a). Extracting a phase from these distributions via ellipse fitting, particularly in the presence of QPN, yields biased results near zero phase or contrast~\cite{foster_method_2002, marti_imaging_2018}. While this means that any useful measurement must operate away from this point, to initially identify an optimal dark time with respect to relative stability we choose to operate in a biased regime with no phase offset. This is because any partitioning that yields a frequency shift results in a phase offset, and thus bias, that varies with dark time, obscuring the optimal value. We characterize this biasing via Monte-Carlo simulations~\cite{Note1} which, when combined with the expected effects of QPN, are in good agreement with the data (Fig.~\ref{fig:atomatom}b).

Guided by these measurements, we perform a 4.3 hour-long synchronous comparison between sub-ensembles at the near-optimal dark time of 15~s. At this long dark time, the diagonal partitioning results in a sufficiently large tweezer-induced phase shift between sub-ensembles to eliminate the effects of biasing (Fig.~\ref{fig:atomatom}bc). This is confirmed both via the same Monte-Carlo simulations used above to characterize bias, and by the agreement between the data and the prediction based on QPN. Specifically, we expect a tweezer-induced frequency offset of 7.0(1.3)~mHz based on previous measurements of the light shift~\cite{shi_polarizabilities_2015, norcia_seconds-scale_2019-1}, and measure an offset of 7.15(18)~mHz. This corresponds to a measurement precision of $4.2\times10^{-19}$. In this unbiased condition, we compute the Allan deviation~\cite{Note1}, which averages down with a slope of $5.2(3)\times10^{-17}~(\tau/s)^{-1/2}$. This is in good agreement with the expected value of $5.2\times10^{-17}~(\tau/s)^{-1/2}$ from QPN with no bias correction (Fig.~\ref{fig:atomatom}c), and comparable to the state of the art value of $3.1\times10^{-17}~(\tau/s)^{-1/2}$ for such synchronous comparisons reported in leading 3D lattice clocks~\cite{campbell_fermi-degenerate_2017-1}. Moreover, the long interrogation times used here allow us to match the highest duty cycles achieved in our previous work of 96\%~\cite{norcia_seconds-scale_2019-1}, even without performing repeated interrogation. As a result, while not demonstrated here, Dick effect noise is not expected to significantly impact the stability of an asynchronous comparison~\cite{norcia_seconds-scale_2019-1}.

\begin{figure}[!t]
    \includegraphics[width=\linewidth]{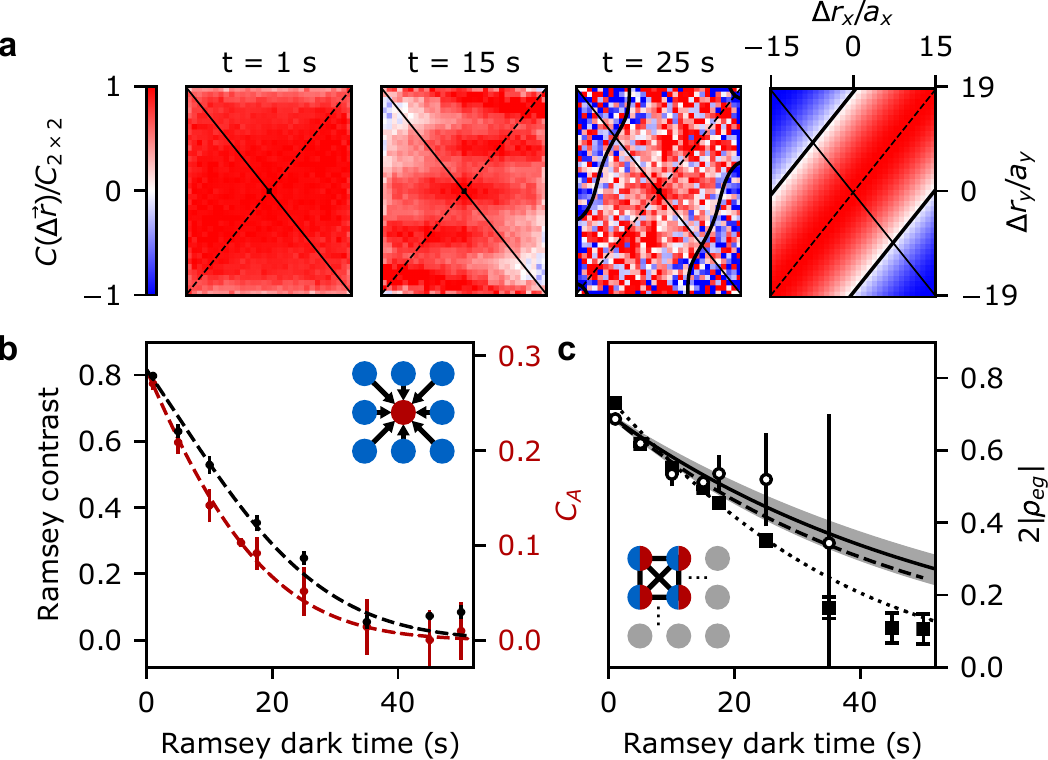}
    \caption{\textbf{Microscopic studies of atomic coherence.} a) As a measure of atomic coherence we compute the spatially resolved atom-atom correlation function, $ C(\Delta\Vec{r}) $~\cite{Note1}, as a function of dark time. These plots are normalized by $C_{2\times2}$, as defined below, to isolate the effects of dephasing from atom loss and decay. The relative displacements $\Delta r_{x,y}$ are normalized by the array spacing in the relevant direction ($a_x = 1.5~\mathrm{\mu m}$, $a_y = 1.2~\mathrm{\mu m}$) such that the pixel spacing corresponds to the tweezer spacing. Solid (dashed) diagonal lines indicate axes along which the tweezer wavelengths change (remain constant) showing accelerated (reduced) dephasing along the forward (reverse) diagonal of the array due to tweezer-induced frequency shifts (solid contour shows where the correlator passes through zero). The fourth frame is a theoretical prediction at 25~s given our known tweezer frequencies and depths. b) The coherence of a single atom (red circle in cartoon) can be measured by computing the average correlations $C_A$ between it and an ensemble of reference atoms (blue circles). In this case the reference ensemble is the entire array, and the excess decay of $C_A$ (red points, averaged over a $4\times4$ block of atoms at the center of the array) compared to the decay of the Ramsey contrast (black points) can be used to quantify the single-atom coherence time. Fits to these quantities (dashed lines) with a Gaussian and exponential component yield overall $1/e$ times of 14.6(7)~s and 19.5(8)~s respectively. c) Based on these measurements, we can infer a single-atom coherence time of 48(8)~s (dashed line)~\cite{Note1}, which is in good agreement with a model based on the measured lifetimes and initial Ramsey contrast (solid line, error in grey). Open circles are $C_A$ with the decay associated with the reference ensemble divided out, which serves as a direct measurement of the single-atom coherence $|\rho_{eg}|$. In the absence of dephasing, an ensemble of these atoms would have a Ramsey contrast of $2|\rho_{eg}|$~\cite{Note1}. To extend this measurement from the central $4\times4$ region to the full array, we consider the average correlation between all atom pairs in a $2\times2$ block averaged over all such blocks, $C_{2\times2}$. In this case each atom in the block acts as a reference for all other atoms in the block (see cartoon). The square root of this quantity (black squares) decays with a fitted $1/e$ time of 33(2)~s (double-dashed line), and serves as a lower bound on the average atomic coherence across the entire array~\cite{Note1}.
    }
    \label{fig:correlation}
\end{figure}

To better understand the limitations of this system, we are interested in characterizing atomic coherence in the absence of tweezer-induced dephasing. To do this, we look for classical correlations in the states of the atoms after Ramsey evolution. Specifically, we compute the $g^{(2)}$ correlator~\cite{Note1} between atoms in different tweezers as a function of Ramsey dark time and relative tweezer position $\Delta\Vec{r}$, which we denote as $C(\Delta\Vec{r})$ (Fig.~\ref{fig:correlation}a)~\cite{chwalla_precision_2007, chou_quantum_2011, hume_probing_2016}. After averaging over the phase of the laser, for two atoms 1 and 2 each with density matrix $\rho_{j=1,2}$, the correlator is equal to $ 2A_1 A_2 \cos(\phi_1-\phi_2)$ where $\rho_{eg,j} = A_j e^{i\phi_j}$, assuming perfect $\pi/2$-pulses in the Ramsey sequence~\cite{Note1}. This quantity correspondingly serves as a site-resolved measure of tweezer-induced clock transition shifts, revealing that along the forward diagonal of the array, where frequency offsets between tweezers --- and thus clock frequency offsets --- are maximal, the atoms become uncorrelated, and eventually develop negative correlations. Along the anti-diagonal, where there is no frequency offset between tweezers, positive correlations persist over much longer timescales. We further observe the development of fringes in the correlator along the more tightly spaced axis of the array, which we hypothesize are the result of overlaps between tweezers~\cite{Note1}.

For a given atom, $\rho_{eg}$ may be defined with respect to a partner atom, or an ensemble of atoms, which serves as a phase reference~\cite{chou_quantum_2011, marti_imaging_2018, tan_suppressing_2019-1}. If the atom and reference are at the same frequency, any excess decay of correlations between the atom and reference compared to the decay of the reference can be attributed to loss of single-atom coherence~\cite{Note1}; if the frequencies are different, the signal falls more rapidly due to the evolving phase difference and constitutes a lower bound. With this in mind, we can compare the average correlations between one atom and all other atoms in the array, $C_{A}$, with the measured Ramsey contrast (Fig.~\ref{fig:correlation}b). $C_{A}$ can equivalently be interpreted as the correlator between one atom and the total spin projection of the remaining array. Applying this procedure to the central $4\times4$ sites, which have a clock frequency comparable to that of the array mean, we infer a single-atom $1/e$ coherence time of 48(8)~s and a resulting atomic oscillator quality factor of $Q=6.5(1.1)\times10^{16}$. This can be compared with the expected coherence time without tweezer-induced frequency shifts of 55(8)~s. This coherence time corresponds to the useable timescale for frequency comparison measurements as in Fig.~\ref{fig:atomatom} that we would expect in the absence of tweezer-induced dephasing, as might be achieved with the use of a spatial light modulator.

In order to extend this argument to each atom in the array, particularly to atoms whose clock frequencies differ substantially from the ensemble mean, we use only atoms that have a similar frequency to the atom under measurement as a phase reference. Specifically, we consider $2\times2$ sub-ensembles of the array, for which we expect tweezer-induced dephasing to be suppressed to a timescale of several hundred seconds. In this case, the sub-ensemble-averaged single-atom coherence can be written in terms of the average of the pairwise correlators~\cite{Note1}. With reasonable assumptions~\cite{Note1}, the square root of this quantity averaged across all such sub-ensembles contained in the array, $\sqrt{C_{2\times2}}$, provides a lower bound on the average atomic coherence $|\bar{\rho}_{eg}|$ of all atoms in the array. This bound has a measured $1/e$ lifetime of 33(2)~s, corresponding to greater than half-minute scale coherence between $\sim150$ atoms on an optical transition.

These coherence times and atom numbers have advanced the state of the art in atomic coherence at optical frequencies, and pushed tweezer clocks to a new regime of relative stability. These advancements hinge on our development of a new recipe for creating tailored optical potentials that balance desirable properties in terms of efficiency, control, and preservation of atomic coherence. This is accomplished by interfacing multiple 2D tweezer arrays at different wavelengths with a standing wave optical lattice. The result is a substantial increase in accessible sample sizes to hundreds of tweezers in this work, and a clear path towards scaling to more than a thousand tweezers~\cite{Note1}.

A key limitation of this platform is the relatively high atom loss incurred when imaging in 515~nm potentials~\cite{cooper_alkaline-earth_2018, norcia_microscopic_2018} compared to the performance possible in 813~nm tweezers~\cite{covey_2000-times_2019, norcia_seconds-scale_2019-1}. While this is not a significant issue for clock performance, it is relevant for gate or many-body based protocols for generating entanglement, where state purity can be critical. The imaging fidelity could be addressed by imaging in a deep 813~nm 3D lattice, which can create tightly confining potentials more efficiently than a tweezer array. Such an approach would have the added benefit of improving our Lamb-Dicke parameter for clock spectroscopy, and, correspondingly, the contrast of our Rabi pulses. For the imaging performance \cite{covey_2000-times_2019, norcia_seconds-scale_2019-1} and confinement available in such a potential, fidelities of $99.9\%$ are readily achievable~\cite{Note1}. In this case a 515~nm tweezer array and axial lattice would still be required for performing high fidelity ground-state cooling via the $\mathrm{^1S_0} \Leftrightarrow \mathrm{^3P_1}$ transition, and would further be useful for performing site-resolved rearrangement in the lattice. Indeed, preliminary results of loading from a tweezer array into a 2D lattice potential at 813~nm, already integrated into our apparatus, showed that low temperatures were achievable. 

The advances in this work are, in part, guided by ground-breaking studies in optical lattice clocks~\cite{hutson_engineering_2019}. Our observations might also illuminate new paths forward for lattice systems that benefit from greater atom number than tweezer clocks. While the elimination of tunneling in this work is partially due to increased trap separation in comparison to lattice clocks, a far greater effect is the presence of disorder. Specifically, as is well-known in tweezer systems~\cite{kaufman_two-particle_2014, murmann_two_2015}, tweezer-to-tweezer disorder is hard to suppress on the energy-scale of the tunneling. While this is a challenge for their use in Hubbard physics, here it serves to suppress tunneling and prolong atomic coherence. This suggests that, in the context of lattice clocks, the use of a weak disordering potential super-imposed on a standard optical lattice clock could enhance coherence time, which might be an alternative solution to directly modulating the tunneling~\cite{hutson_engineering_2019}. This highlights another important role for the tweezer clock: it serves as a clean, versatile platform for studying neutral-atom optical clocks and the mechanisms that influence their performance. In future accuracy studies, the lack of interactions and itinerance in this system will ease dissection of coupled systematic effects. 

Our work here lays a firm foundation for implementation of entanglement on an optical clock transition~\cite{gil_spin_2014}. The combination of large, tightly spaced 2D ensembles with long-lived atomic coherence is the ideal starting point for engineering interactions via Rydberg excitations driven from a long-lived excited state~\cite{gil_spin_2014, wilson_trapped_2019, madjarov_high-fidelity_2020-1}. This opens up several exciting possibilities, including the creation of metrologically useful entangled states like squeezed~\cite{gil_spin_2014, kaubruegger_variational_2019} or GHZ states~\cite{omran_generation_2019}, probing and verifying the resulting entanglement with microscopic observables, and, in the context of quantum simulation, implementing various 2D spin models of interest~\cite{zhang_topological_2015, savary_quantum_2017, titum_probing_2019}. For applications in quantum information, such a system can also be used to perform Rydberg-mediated quantum gates on long-lived spin or optical qubits~\cite{levine_parallel_2019, graham_rydberg-mediated_2019, madjarov_high-fidelity_2020-1}, or to prepare cluster states in a highly parallelized way for use in measurement-based quantum computing~\cite{briegel_measurement-based_2009}.

\droptocpage

%

\section{Methods}

\subsection{Apparatus}

Our procedure for loading, ground-state cooling, and imaging bosonic strontium-88 ($^{88}$Sr) atoms in 515~nm optical tweezers is described in \cite{norcia_microscopic_2018}. Power hungry operations like initial loading and imaging are performed exclusively in these tweezers. We have observed that loading can be performed in even shallower tweezers with the aid of the axial lattice; however, this results in an additional background of atoms that populate other layers of the lattice. To avoid this, we opt to load directly into the tweezers to ensure loading of a single atom plane. The lattice is then ramped on for sideband cooling~\cite{monroe_resolved-sideband_1995, kaufman_cooling_2012, cooper_alkaline-earth_2018, norcia_microscopic_2018} to allow for high-fidelity cooling in the axial direction.

\subsection{Tweezer Arrays}

To prepare our 2D tweezer arrays, we image two orthogonal acousto-optic deflectors (AODs) onto each other in a $4f$ configuration. Two such systems at 515~nm and 813~nm are combined on a dichroic and projected via the same high-numerical-aperture objective lens (NA $>$ 0.65), which has diffraction-limited performance between 461~nm and 950~nm. Relevant 515~nm and 813~nm tweezer parameters are collected in table \ref{table:S1}.

We space the two axes of our array differently, with 1.5 and 1.2~$\mu$m spacings along the two orthogonal axes of the array, corresponding to $\sim 5$~MHz ($\sim 3$~MHz) offsets between adjacent 515~nm (813~nm) tweezers. This keeps nearby tweezers at different optical frequencies, such that any interference is time-averaged away and can be compensated for by trap balancing. For equally spaced tweezers, we have observed DC interference fringes that cannot be removed due to a lack of access to the appropriate degrees of freedom in trap balancing.

\begin{table}[h!]
\centering
\begin{tabular}{|c|c|c|c|} 
 \hline
  & 515~nm & 813~nm \\ [0.5ex] 
 \hline
 Available optical power & ~10 W & ~3 W \\
 Ground-state polarizability, $\alpha^{E1}$ & 900 $a_0^3$ \cite{safronova_extracting_2015} & 280 $a_0^3$ \cite{shi_polarizabilities_2015} \\
 Tweezer $1/e^2$ Gaussian radius & 480(20)~nm & 740(40)~nm \\ [1ex] 
 \hline
\end{tabular}
\caption{Relevant optical trapping parameters for both tweezer systems. The higher polarizability and available laser power, as well as tighter spatial confinement, make the 515~nm tweezers more appropriate for cooling and imaging atoms in larger tweezer arrays, as these operations require more strongly confining traps.}
\label{table:S1}
\end{table}

In order to balance the depths of individual tweezers, we split off a small fraction of the light before the objective, and measure the integrated intensity per tweezer using a CMOS camera. By adjusting the relative power in different RF tones applied to the crossed AODs, it is possible to balance the total optical power in each spot to within 5\% of the mean. The main limitation on this balancing is a lack of fully independent control over each spot: we only have $16+20=36$ degrees of freedom for balancing a tweezer array of $16\times20=320$ spots.

\subsection{Tweezer RF source}

To supply the AODs used to generate our tweezers with appropriate RF signals, we use a custom FPGA-based frequency synthesizer. Specifically, the FPGA runs 512 DDS cores, which are interleaved to generate 256 outputs with independently tunable frequency, phase, and amplitude. These outputs control 4 separate 16-bit digital-analogue converters (DACs) which each drive one of the four AODs used in our system. This corresponds to 64 independent RF tones per AOD, where each tone has 36-bits of frequency resolution, 12-bits of phase resolution, and 10-bits of amplitude resolution. The outputs are clocked at 750~MHz (but can be clocked in the gigahertz range if desired), corresponding to a maximum usable frequency of $\sim300$~MHz (for this work we operate in the 100-200~MHz range). These outputs are amplified using two stages of linear RF amplifiers, with the final stage being a high power (10~W) amplifier that delivers $\sim2$~W ($\sim5$~W) of total RF power to each of the 515~nm (813~nm) AODs.

\subsection{Axial lattice}

To form the axial lattice, $\sim$300 mW of 515~nm light is split in an interferometer that creates two parallel beams with variable spacing and controllable relative phase. These two beams are focused onto the atoms with a 30~mm achromatic doublet, such that they form a standing wave with k-vector normal to the tweezer plane. For the chosen beam spacing of 1.6~cm at the lens, the resulting lattice potential has a period of $\lambda_l \approx 1~\mu$m. Each lattice beam has a Gaussian $1/e^2$ radius of 25~$\mu$m at the atoms.

\subsection{Clock path}

Our 698~nm clock light comes from a laser injection locked with light stabilized to a cryogenic silicon reference cavity \cite{oelker_demonstration_2019}. The output of this injection lock travels through a 50 m long noise-cancelled fiber.

For the Rabi spectroscopy presented in this work, the clock path further included $\sim4$~m of fiber and $\sim$50~cm of free-space path which were un-cancelled and added phase noise to our clock light. 

For all remaining data, phase noise cancellation was performed using a reference mirror attached to the objective mount which, to first order, sets the position of the tweezer array. This left only $\sim2$~m of un-cancelled fiber in the path, but did not noticeably improve the atom-light coherence of the system.

\subsection{Repumping}

Our clock-state lifetime measurements can be confounded by the presence of atoms pumped into the $\mathrm{^3P_2}$ state due to Raman scattering of the trap light. These atoms are not distinguished from clock-state atoms during our normal blow-away measurement, and can lead to an artificially long inferred lifetime. To avoid this, we add an additional repumping step that depletes $^3$P$_2$ atoms before the blow-away by driving the $^3$P$_2$ $\Leftrightarrow$ $^3$S$_1$ transition at 707~nm. Note that since $^3$S$_1$ decays to $^3$P$_0$ with a branching ratio of $ \sim1/9 $, this measurement alone is insufficient to accurately determine the population in $^3$P$_0$. As a result, we repeat the above measurement without repumping to measure the total $^3$P$_0+^3$P$_2$ population. Based on these two measurements we infer the true population in $^3$P$_0$, which appears in Fig.~\ref{fig:lifetime}a.

\subsection{Units}

Throughout this article and its supplement, unless otherwise stated, when we quote a lifetime we are referring to the $1/e$ decay time. As a frequency, the inverse of this quantity may be read as radians per second.

\section{Data availability}

The experimental data presented in this manuscript is available from the corresponding author upon reasonable request.

\section{Acknowledgements}

We acknowledge fruitful discussions with R. B. Hutson, J. K. Thompson, M. Foss-Feig, S. Kolkowitz, and J. Simon. We further acknowledge F. Vietmeyer and M. O. Brown for invaluable assistance in the design and development of our FPGA-based tweezer control system. This work was supported by ARO, AFOSR, DARPA, the National Science Foundation Physics Frontier Center at JILA (1734006), and NIST. M.A.N., E.O., and N.S. acknowledge support from the NRC research associateship program.

\section{Author contributions}
A.W.Y., W.J.E., M.A.N., N.S., and A.M.K. built and operated the tweezer apparatus, and the silicon-crystal stabilized clock laser was operated by W.R.M., D.K., E.O. and J.Y. All authors contributed to data analysis and development of the manuscript. 

\section{Author information}
The authors declare no competing interests.



\clearpage
\widetext
\appendix

\setcounter{section}{0}
\setcounter{equation}{0}
\setcounter{figure}{0}
\setcounter{table}{0}
\setcounter{page}{1}
\makeatletter
\renewcommand{\theequation}{S\arabic{section}.\arabic{equation}}
\renewcommand{\thefigure}{S\arabic{figure}}
\renewcommand{\thetable}{S\arabic{table}}

\renewcommand{\tocname}{Supplemental Materials}
\renewcommand{\appendixname}{Supplement}

\clearpage

\tableofcontents
\appendix
\setcounter{secnumdepth}{2}

\incltocpage
\section{Alignment of optical potentials}

\subsection{Tweezer array alignment}

The two tweezer arrays share a common reference - the microscope objective used to project them - and so alignment of the tweezers is both fairly straightforward and robust, with negligible drifts on the 0.5~$\mu$m scale over the course of multiple days. To initially overlap the two tweezer arrays we align the imaging system to the 515~nm tweezers, and take destructive fluorescence images of atoms in the 813~nm tweezers. Varying the focus of the 813~nm system to bring the atoms into focus aligns the arrays in the axial direction. For fine alignment of the in-plane position of the 515~nm and 813~nm tweezer arrays, we scan the RF tones used to generate the 813~nm tweezers, optimizing for low atom loss and temperature after passing atoms between the two tweezer arrays.

\subsection{Axial lattice alignment}

\begin{figure}[!htb]
    \includegraphics[width=130mm]{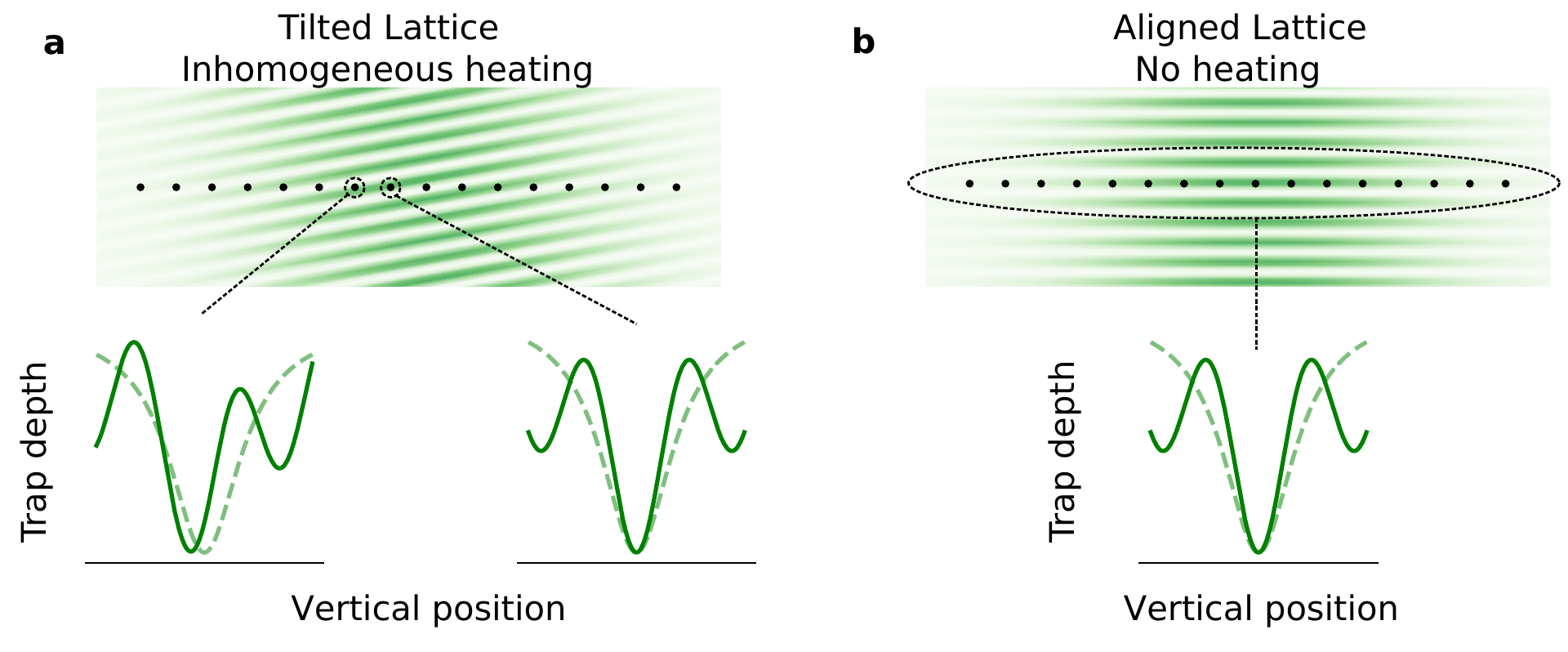}
    \caption{Axial lattice alignment via parametric heating. a) The upper panel shows a side view of a misaligned axial lattice (green fringes) overlapped with the tweezer array (black dots). The lower plots show the tweezer potential (dashed curve) and the combined lattice and tweezer potential (solid green curve) at two locations. When the lattice fringe is tilted relative to the plane of the tweezers, it shifts the centers of the traps (as shown in the bottom left plot) in a position-dependent way. As such, modulating the depth of the tweezers will cause different parametric heating rates at different sites in the array. b) When the tweezers and lattice are well aligned, modulating the tweezers at the trap frequency ($\omega_m = \omega_0$) does not cause substantial heating.}
    \label{sfig:lattice_alignment}
\end{figure}

The axial lattice is projected with an independent lens oriented perpendicular to the objective, and so does not share a convenient common reference with the tweezers. As such, both the position and angle of the fringes relative to the tweezer plane must be carefully optimized. The lattice beams are initially overlapped with the tweezers by loading atoms directly from the MOT into the lattice. Fluorescence images of these atoms are then used to center the lattice on the tweezer array. The axial alignment is coarsely optimized by measuring light shifts induced by the lattice on the $^1\rm{S}_0 \Leftrightarrow ^3\rm{P}_1$ transition in non-magic magnetic fields.

To flatten the lattice fringes relative to the tweezers we parametrically heat atoms trapped in the axial lattice with the tweezers. Specifically, in the case that a tweezer is well centered on a lattice fringe, modulating the power in the lattice changes the overall trap depth experienced by an atom, but not the position of the trap center (Fig.~\ref{sfig:lattice_alignment}b). This results in parametric heating with a resonance at $\omega_m = 2\omega_0$, where $\omega_0$ is the trap frequency in the absence of modulation~\cite{savard_laser-noise-induced_1997, jauregui_nonperturbative_2001}. In this case, if the modulation frequency is equal to the trap frequency, the atom is not strongly heated. If, on the other hand, the lattice and tweezer are misaligned, this modulation results in a shaking of the trap center, which can result in heating at $\omega_m = \omega_0$ (Fig.~\ref{sfig:lattice_alignment}a)~\cite{savard_laser-noise-induced_1997, jauregui_nonperturbative_2001}. For an appropriately chosen modulation amplitude and duration at $\omega_m = \omega_0$, we can use the observed probability of loss in a given tweezer as an indication of the relative alignment between that tweezer and the nearest lattice fringe. By scanning the phase of the lattice and fitting the phase of the resultant heating signal at each tweezer, we extract the relative orientation of the lattice and tweezers (as shown in Fig.~\ref{fig:setup}d of the main text). This signal allows for alignment of the lattice to the tweezers at the $\lambda_l/10$ level, where $\lambda_l \approx 1~\mu$m is the lattice spacing.

Once optimized, this flattening has not been observed to change over multiple weeks. However, the spatial phase of the lattice can drift by a period over several hours due to a combination of thermal expansion and slightly mismatched path lengths in the lattice beams. This could eventually be addressed by actively feeding back on the lattice phase as measured by a camera, or by improving the passive stability of the lattice in a future design.

\section{Atom-light coherence}

\begin{figure}[!htb]
    \includegraphics[width=0.45\linewidth]{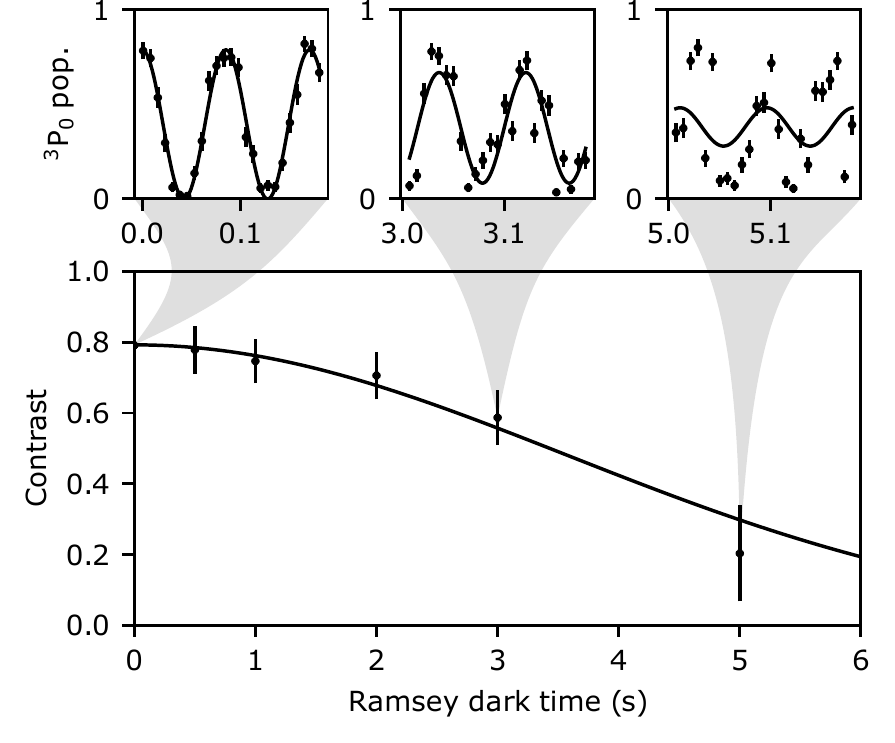}
    \caption{Measuring atom-light coherence. Fitting measured Ramsey fringes with fringes of a fixed frequency provides a conservative estimate of atom-light coherence. Callouts share x-axis units with the main plot, and show the fitted Ramsey data (which is the same data as used in Fig.~\ref{fig:lifetime}b in main text)}
    \label{fig:atom_light_sup}
\end{figure}

As a conservative measurement of our atom-light coherence, we fit the measured Ramsey fringes with frequency as a fixed parameter, which yields a Gaussian lifetime of 3.6(2)~s (Fig.~\ref{fig:atom_light_sup}). This is consistent with the value of 3.4(4)~s measured in our previous work \cite{norcia_seconds-scale_2019-1}.

\section{Experimental sequence}

\begin{figure}[!htb]
    \includegraphics[width=117mm]{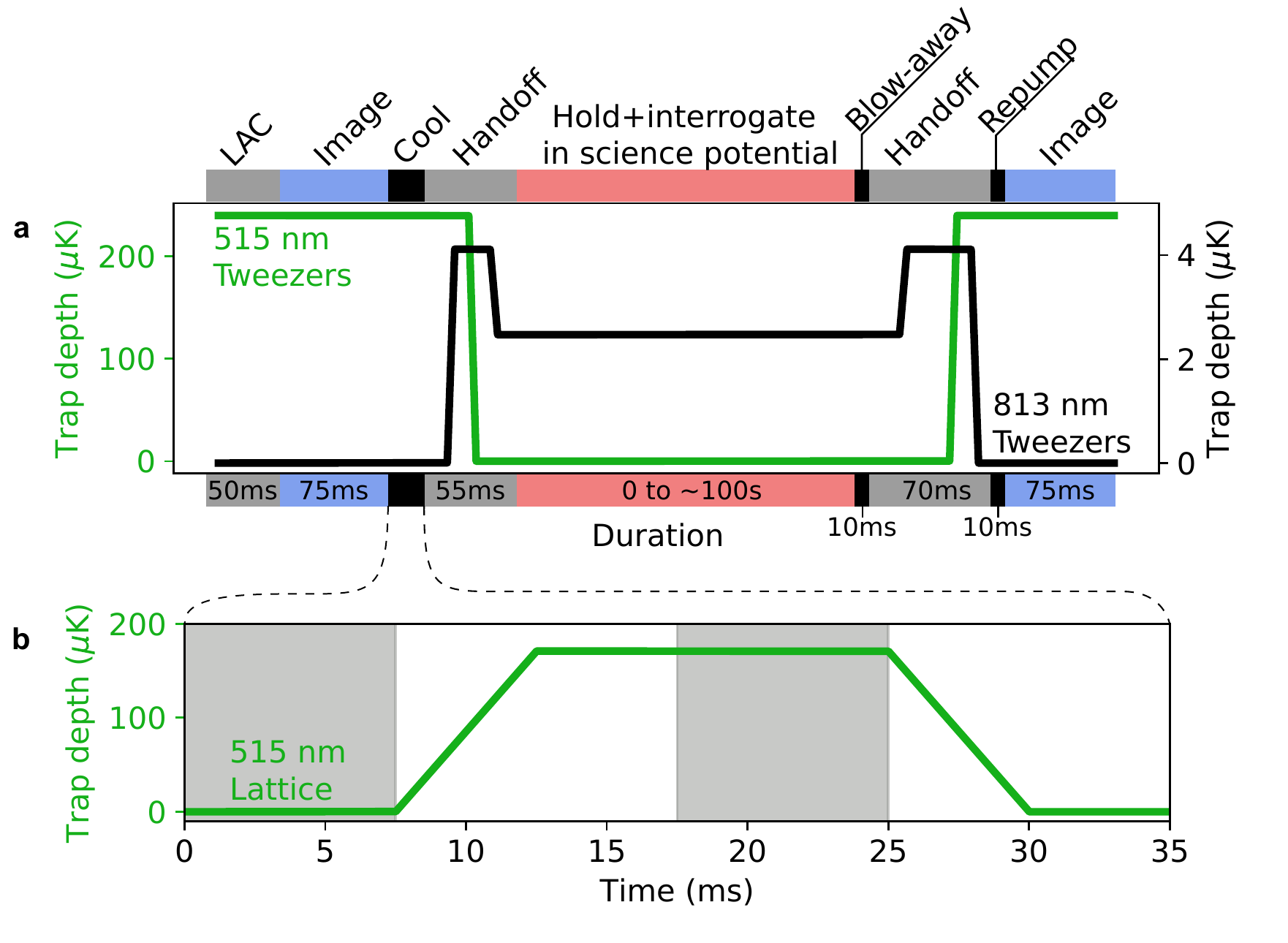}
    \caption{Timing of experimental sequence. a) The green and black curves track the depths of the 515~nm and 813~nm tweezers, respectively. The colored regions above and below this graph categorize each step of the experiment, which are described in more detail in the text.  We find that maintaining the 813~nm tweezers at a depth greater than 20$\mathrm{E_R}$ during the ramp down improves the fidelity of the handoff procedure. Not shown is the time required to load atoms into the 515~nm tweezers from the MOTs, which takes roughly 120~ms. b) A zoomed in view of our cooling procedure, showing the depth of the axial lattice. We perform two rounds of sideband cooling, indicated by the two regions shaded in grey. The first, done before ramping up the axial lattice, does not cool axial motion to the ground state. Instead, it is important for reducing the size of the atomic wave packet to ensure loading of a single lattice fringe.}
    \label{sfig:timing}
\end{figure}

As in our previous works~\cite{norcia_microscopic_2018, norcia_seconds-scale_2019-1}, we load our tweezers from a narrow-line MOT operating on the $^1$S$_0$ $\Leftrightarrow$ $^3$P$_1$ cooling transition, and remove pairs of atoms via light assisted collisions (LACs). The result is random $\sim45$\% filling of the tweezer array with single atoms. Detection is performed via fluorescence imaging on the 461~nm $^1$S$_0$ $\Leftrightarrow$ $^1$P$_1$ cycling transition, while simultaneously performing resolved sideband cooling on the 7.5~kHz wide $^1$S$_0$ $\Leftrightarrow$ $^3$P$_1$ intercombination line to mitigate the effects of recoil heating. For this work, the axial lattice is used primarily for improved 3D ground-state cooling. It is ramped on and off over 5~ms, and is shuttered for all stages of the experiment except for ground-state cooling. To shrink the size of the atomic wave packet and prevent loading of adjacent lattice fringes, we perform 5~ms of un-resolved axial and resolved radial sideband cooling in the tweezers before ramping on the axial lattice. The improved axial confinement with the lattice on creates nearly isotropic traps with $ \sim90 $~kHz trap frequencies along all axes, and further cooling in this hybrid potential brings most of the atoms ($ 81^{+17}_{-22}\% $) to the 3D motional ground state. Since the polarization of the axial lattice is aligned to that of the 515 tweezers, we maintain the same ``magic field'' conditions throughout this sequence~\cite{norcia_microscopic_2018}.

To hand atoms between the two sets of tweezers, we ramp on the 813~nm array in 5~ms with the 515~nm traps maintained at full depth, and then ramp the 515~nm tweezers off. The intensity servo for our 515~nm tweezers takes a few milliseconds to stabilize after being turned back on, which can heat atoms out of the 813~nm tweezers. To avoid this, while the 515~nm tweezers are nominally switched off we also move them away from the atoms with the AODs used to project them, and then shutter the 515~nm beam path. To reintroduce the 515~nm tweezers, we turn on the beam and let the intensity servo settle at low power, un-shutter the beam path, and finally move the tweezers back to overlap with the 813~nm array. By shuttering the beam, we ensure that there are no light shifts of the clock transition due to stray 515~nm light while the atoms are in the 813~nm tweezers. Note that while the handoff procedure can be performed with 0.0(3)\% atom loss, for all clock data in the main text this alignment was imperfect, resulting in an additional $\sim4\%$ atom loss when handing atoms to and back from the science potential. In this work, we choose not to correct this loss because it is inconsequential for clock performance, however, more careful and consistent calibration will be necessary for future works that are more sensitive to state purity.

To probe the ultranarrow $^1$S$_0$ $\Leftrightarrow$ $^3$P$_0$ clock transition, we apply 698~nm clock light 125~kHz off resonance. Once the laser intensity servo settles, we can jump the detuning to near resonance for a variable time to excite atoms to the $^3$P$_0$ state. The population in this state is then measured by using 461~nm light (resonant with $^1$S$_0$ $\Leftrightarrow$ $^1$P$_1$) to heat --- or ``blow-away'' --- ground-state ($^1$S$_0$) atoms in the tweezers. To return clock-state atoms to the ground state for readout, we drive the  $^3$P$_0$ $\Leftrightarrow$ $^3$S$_1$ transition at 679~nm and the $^3$P$_2$ $\Leftrightarrow$ $^3$S$_1$ transition at 707~nm. The $^3$S$_1$ state decays to the whole $^3$P$_J$ manifold, such that eventually all clock-state atoms are pumped into the shorter-lived $^3$P$_1$ state and decay back to the ground state, where they can be read out during imaging. For Ramsey spectroscopy, we use a $\pi/2$-pulse time of $\sim50$~ms for all relevant data in this paper.

\section{Lifetime limits}

\subsection{Deep traps}\label{sec:deepTraps}

We expect our trap lifetimes, particularly in deeper traps, to be limited by collisions with residual background gas. These collisions are substantially more energetic than the trap depths we have access to, resulting in a vacuum lifetime that is effectively independent of trap depth~\cite{bali_quantum-diffractive_1999}. This is confirmed via the procedure in \cite{van_dongen_trap-depth_2011}, assuming that the main collision partners are room temperature $\Sigma$ state $\mathrm{H_2}$ molecules interacting via Van der Waals forces.

Based on this model, we expect clock-state atoms to have reduced trap lifetimes $\tau$ compared to ground-state atoms due to their larger $C_6$ coefficient and thus larger scattering cross section $\sigma$, since in this case $\tau \propto 1/\sigma \propto C_6^{-2/5}$. With the known $C_6$ coefficients for collisions between $\mathrm{H_2}$ and $\mathrm{^{88}Sr}$ \cite{mitroy_dispersion_2010} we calculate that the ratio between the ground and clock-state trap lifetime ($\tau_g$ and $\tau_e$ respectively) is $\tau_{g}/\tau_{e} = 1.10$, which agrees with results from \cite{dorscher_lattice-induced_2018}.

We estimate that the fractional frequency shifts due to these background collisions \cite{gibble_scattering_2013, bothwell_textjila_2019} are below the $ 10^{-19} $ level, suggesting that this is not a likely explanation for the increased decoherence rate observed in the main text.

\subsection{Shallow traps}

The source of the dramatic reduction of trap lifetime in shallow traps is as of yet unknown; however, based on the above analysis we rule out the effect of collisions with background gas. Other potential sources could include tunneling, or heating induced by parametric modulation, pointing noise, or scattered light.
\droptocpage
\subsubsection*{Tunneling}

For 6$\mathrm{E_R}$ deep tweezers with $1.2~\mathrm{\mu m}$ spacing, we calculate a tunneling rate of $\sim1$~Hz between adjacent tweezers via exact diagonalization in 1D. For image pairs in the experiment, we expect this tunneling to manifest as correlated nearest-neighbor atom-vacancy and vacancy-atom events, where an atom tunnels from one site to an empty adjacent site, or pairs of atom-vacancy events, where an atom tunnels onto an occupied adjacent site, and both atoms are lost due to light-assisted collisions. We do not observe an excess of such events beyond what is expected due to loss and imaging infidelity at any depth or hold time used in this work. This suggests that disorder, which we know is present on the scale of $>10^{-2}$ in trap depth, plays a critical role in pinning the atoms. Given the relevant tunneling energies this is not unexpected, since even in our shallowest traps disorder on the scale of $10^{-4}$ in trap depth is sufficient to freeze out tunneling. Similar calculations suggest that the effect of loss due to Zener tunneling along the gravitational axis is negligible at all depths explored in this work.

\droptocpage
\subsubsection*{Heating}

Optically trapped atoms can be heated through a variety of mechanisms, including intensity and pointing noise from the trap laser, and scattered light. Intensity noise manifests as parametric modulation of the trap frequency which, assuming a flat noise spectrum, results in exponential heating (measured in phonon number) with a time constant proportional to $f_{t}^2$, where $f_t$ is the trap frequency. Similarly, pointing noise with a flat spectrum results in linear heating with a rate proportional to $f_t^3$~\cite{savard_laser-noise-induced_1997, gehm_dynamics_1998}. For comparison, the number of bound states $N$ in a tweezer scales roughly like $\sqrt{U}\propto f_t$, where $U$ is the trap depth. As such, assuming a flat noise spectrum, both these sources of loss should improve with reduced trap depth.

While the intensity noise of our trapping laser is suppressed below 10~kHz via a servo, and otherwise relatively flat over the frequencies relevant for heating, we do not expect this to be true for pointing noise. In this case there is increased noise at lower frequencies due to mechanical resonances and acoustic noise, and no convenient way of removing such noise with a servo. As a result pointing noise likely contributes to our reduced lifetimes at and below trap depths of 15$\mathrm{E_R}$ (corresponding to 6.8~kHz trap frequencies). Other sources of trap-independent heating, like scattered background light, can also begin to dominate as the traps become very shallow and $N$ becomes small.

\incltocpage
\subsection{Clock-state lifetime and coherence time}\label{sec:clockLifetime}

\begin{table}[h!]
\centering
\begin{tabular}{|c|l|} 
 \hline
 \multicolumn{2}{|c|}{Values inferred from measurement (s)}\\
 \hline
 $ 1/\Gamma^t_g $ & 101(6)\\
 $ 1/\Gamma^t_e $ & 92(5)$^*$\\
 $ 1/\Gamma_e $ & 43(4)\\
 \hline
 \multicolumn{2}{|c|}{Theory values ($\rm \times10^{-3}~s^{-1}$)}\\
 \hline
 $ \Gamma^\mathrm{BBR}_0~(\rm ^3P_0 \Rightarrow \,^3P_0) $ & $3.45(22)$~\cite{dorscher_lattice-induced_2018}\\
 $ \Gamma^\mathrm{BBR}_1~(\rm ^3P_0 \Rightarrow \,^3P_1) $ & $2.23(14)$~\cite{dorscher_lattice-induced_2018}\\
 $ \Gamma^\mathrm{BBR}_2~(\rm ^3P_0 \Rightarrow \,^3P_2) $ & $0.105(7)$~\cite{dorscher_lattice-induced_2018}\\
 $ \Gamma^\mathrm{R}_0~(\rm ^3P_0 \Rightarrow \,^3P_0) $ & $0.557~\rm (U/E_R)$~\cite{dorscher_lattice-induced_2018}\\
 $ \Gamma^\mathrm{R}_{12}~(\rm ^3P_0 \Rightarrow \,^3P_1, \,^3P_2) $ & $0.782~\rm (U/E_R)$~\cite{dorscher_lattice-induced_2018}\\

 \hline
\end{tabular}
\caption{Measured and theoretical values contributing to predicted Ramsey lifetime. All measured values are for a trap depth of $\rm 15~E_R$, based on interpolating between the nearest points in Fig.~\ref{fig:lifetime}a of the main text. * indicates that the inferred value of $\Gamma^t_e$ is dependent on the reasoning and theory values presented in section \ref{sec:deepTraps}. Note that $ \Gamma^\mathrm{BBR}_2 $ is smaller than the error bars on the other processes, and so we neglect this process in our analysis. $\rm U$ is the trap depth and $\rm E_R$ is the recoil energy of a magic 813 nm photon. }
\label{table:rates}
\end{table}

The $ \mathrm{^3P_0} $ clock-state lifetime is primarily limited by the loss processes described above, as well as by scattering of black-body radiation and the trapping light. Because we directly measure the $\mathrm{^3P_0}$ population in the main text, we are sensitive to all processes that remove population from this state, including transitions to the metastable $\mathrm{^3P_2}$ state. Raman scattering of the trap light can drive such transitions, with dominant contributions from $ \mathrm{^3P_0}\rightarrow\mathrm{^3P_1} $ and $ \mathrm{^3P_0}\rightarrow\mathrm{^3P_2} $. For $\pi$-polarized trap light, these processes occur with rates of $\mathrm{4.98\times10^{-4}~s^{-1}E_R^{-1}}$ and $\mathrm{2.84\times10^{-4}~s^{-1}E_R^{-1}} $ respectively~\cite{dorscher_lattice-induced_2018}. Note that while the ratio of these two scattering processes is polarization-dependent, their sum, with a value of $ \Gamma^{\rm R}_{12} = 7.82~\rm \times10^{-4}~s^{-1}E_R^{-1}$, is conserved. All population driven into $\mathrm{^3P_1} $ can be assumed to immediately decay into the ground state, whereas processes that return population in $\mathrm{^3P_2} $ to the $ \mathrm{^3P_0} $ state are negligible. As such, to good approximation, $ \Gamma^{\rm R}_{12} $ can be treated as the total rate at which population in $\mathrm{^3P_0}$ is depleted due to Raman scattering.

Black-body radiation can off-resonantly drive transitions to the $\mathrm{^3D_1}$ state, which decays to the $\mathrm{^3P_J}$ manifold with branching ratios of $ R^D_J = $~59.65\%, 38.52\%, and 1.82\% for the $ J=0, 1 $ and 2 states respectively~\cite{dorscher_lattice-induced_2018}. The dominant mechanism by which black-body radiation contributes to decay of the $\mathrm{^3P_0}$ state is via population that branches from $\mathrm{^3D_1}$ into $\mathrm{^3P_1}$, and subsequently decays into the ground state. This process occurs at a rate of $\mathrm{2.23\times10^{-3}~s^{-1}}$~\cite{dorscher_lattice-induced_2018} at room temperature, which we will call $ \Gamma^{\rm BBR}_1 $. The sum of these effects with the rate at which $\mathrm{^3P_0}$ state atoms are lost from the tweezers, $ \Gamma^t_e $, is in good agreement with the our measured $ \mathrm{^3P_0} $ decay rate, $ \Gamma_e = \Gamma^t_e + \Gamma^{\rm R}_{12} + \Gamma^{\rm BBR}_1 $ (theory curve in Fig.~\ref{fig:lifetime}a of the main text).

We can compute an expected Ramsey coherence time given these decay rates. Due to the magic-wavelength traps, Rayleigh scattering of the trap light does not cause decoherence~\cite{dorscher_lattice-induced_2018}. As a result, trap-induced scattering only contributes to decay of Ramsey contrast through the Raman scattering processes described above that remove population from the $ \mathrm{^3P_0} $ state. Unlike Rayleigh scattering of the trap light, black-body processes that drive population out of and back into the $\mathrm{^3P_0}$ state (predominantly via $\mathrm{^3D_1}$) can serve as an extra source of decoherence that is not directly reflected in the $\mathrm{^3P_0}$ lifetime measurement. Including all these effects, the inferred Ramsey coherence time is:

\begin{equation}
	\tau = 2/(\Gamma^t_e + \Gamma^t_g + \Gamma^{\rm R}_{12} + \Gamma^{\rm BBR}_{1}(1 + \frac{R_0}{R_1})),
\end{equation}

\noindent where $ \Gamma^t_g $ is the ground-state atom loss rate. All of these relevant rates are summarized in table~\ref{table:rates}. Note that due to the use of $\mathrm{^{88}Sr}$ in this work, and given the strength of the magnetic fields used, the effects of spontaneous emission from the clock state are negligible in this analysis.

\section{Model for light shifts in an optical tweezer potential}\label{sec:lightshiftDerivation}

The geometry of an optical potential introduces important corrections to standard first order expressions for AC Stark shifts of atomic transitions. In this section, we follow the approach taken in~\cite{ovsiannikov_multipole_2013-2}, which studies light shifts in lattice potentials. We derive an analogous expression for optical tweezers (see also~\cite{madjarov_atomic-array_2019}), and isolate terms that are first-order-sensitive to the detuning of the trapping wavelength from the magic wavelength. 

The intensity profile of an optical tweezer in cylindrical coordinates ${(z,~\rho = \sqrt{x^2 + y^2})}$ is $I(\rho,z) = I_0 \bigl{[} w_0/w(z) \bigr{]}^2 e^{-2\rho^2 / w(z)^2}$, where $I_0$ is the peak intensity of the tweezer, and $w_0$ is the Gaussian beam waist at the focus. The waist as a function of the propagation distance $z$ is $w(z) = \sqrt{1 + (z/z_R)^2}$, where $z_R = \pi w_0^2/\lambda$ is the Rayleigh range. The optical potential is then $U(\rho,z) = \alpha^{E1}(\nu_T) I(\rho,z) + \beta(\nu_T) I(\rho,z)^2$, where $\alpha^{E1}(\nu_T)$ and $\beta(\nu_T)$ are the electric-dipole  polarizability and hyperpolarizability, respectively, of $^{88}$Sr trapped in a tweezer with optical frequency $\nu_T$. The effects of magnetic-dipole and quadrupole polarizabilities in $^{88}$Sr are not included in this analysis.

After expanding the potential $U(\rho,z)$ to fourth order in both $\rho$ and $z$, we treat the quartic corrections as a perturbation to the harmonic solutions, and arrive at an approximate expression for the energy shift $\Delta E_{g,e}$ for the ground ($g$) and excited ($e$) electronic states. The resulting frequency shift $\nu_{LS}$ is expressed $\nu_{LS} = (\Delta E_e -  \Delta E_g)/h$, where $h$ is Planck's constant. For this work, we are interested in how $\nu_{LS}$ varies as a function of $\nu_T$, as the inhomogeneities in $\nu_T$ across our tweezer array serve as a dominant source of dephasing between atoms. 

In order to express our result in terms of the differential electric-dipole polarizability $\frac{1}{ \alpha_{\text{m}}^{E1}} \frac{ \partial  \Delta \alpha^{E1}}{\partial \nu}$, where $ \alpha_{\text{m}}^{E1} $ is the polarizability at the magic trapping frequency $\nu_{\rm m} \simeq 369$~THz, and $ \Delta \alpha^{E1}$ is the difference between the electric-dipole polarizabilities of the ground and excited states, we expand our result in powers of $\sqrt{I_0}$ up to $I_0^2$. Including only terms that depend explicitly on the detuning of the tweezer frequency from the clock-magic trap frequency, $\delta_L = \nu_T - \nu_{\rm m}$, the resulting equation for the differential light shift is

\begin{align}
\frac{h \nu_{LS}}{\rm{E_R}} &\approx \Bigl{[}  \Bigl{(} \sqrt{2} \frac{w_0}{z_R} (n_x + n_y + 1) + \bigl{(}\frac{w_0}{z_R}\bigr{)}^2 (n_z + \frac{1}{2} ) \Bigr{)} (\sqrt{u_e} - \sqrt{u_g}) -  (u_e - u_g) \Bigr{]} \label{eq:LS1}\\
&\approx \Bigl{[}  \frac{1}{2}\Bigl{(} \sqrt{2} \frac{w_0}{z_R} (n_x + n_y + 1) + \bigl{(}\frac{w_0}{z_R}\bigr{)}^2 (n_z + \frac{1}{2} ) \Bigr{)} \sqrt{u_{\text{m}}} -  u_{\text{m}} \Bigr{]} \frac{1}{\alpha_{\text{m}}^{E1}} \frac{\partial \Delta \alpha^{E1}}{\partial \nu} \delta_L \label{eq:LS2}. 
\end{align}
In this expression, $u_{g,e} = \alpha_{g,e}^{E1}(\nu_T) I_0/\rm{E_R}$ which, in going from Eqn. \ref{eq:LS1} to Eqn. \ref{eq:LS2}, are Taylor expanded around $u_{\rm{m}} = \alpha_{\rm{m}}^{E1} I_0/\rm{E_R}$. Finally, $n_z$, $n_x$, and $n_y$ are the number of motional quanta in the axial direction, and two radial directions, respectively. Using the values in table~\ref{table:S1} of the methods section, as well as $\frac{1}{ \alpha_{\text{m}}^{E1}} \frac{ \partial  \Delta \alpha^{E1}}{\partial \nu} = -15.5(1.1)~\rm{(\mu Hz/E_R)/MHz}$~\cite{middelmann_high_2012, shi_polarizabilities_2015}, this calculation yields results that are within 3\% of the values measured via ellipse fitting. Note that because there is substantially more than 3\% error in the characterization of our trap depths and shape, we opt to use the measured values of these shifts to predict the expected Gaussian decay rate associated with dephasing which appears in the main text (see Fig.~\ref{fig:lifetime}b).

\section{Ellipse fitting} \label{sec:ellipse}

To extract the relative phase between two sub-ensembles in the array after a Ramsey measurement, we parametrically plot the excitation fraction in one sub-ensemble against the other. Fitting an ellipse to this distribution via its singular value decomposition, we extract values for the semi-major and minor axes of the best-fit ellipse, $a$, $b$. These are related to the relative phase between ensembles by $\phi = 2\arctan(b/a)$.

We follow the procedure in \cite{marti_imaging_2018} to calculate an expected uncertainty in $\phi$ due to QPN, with a slight modification to account for loss of Ramsey contrast due to atom loss as opposed to loss of coherence. Specifically, the excitation probability in the two ensembles are:

\begin{align}\label{pe}
p_x &= \frac{c_l}{2}\left(1+c_d \cos\theta\right)\\
p_y &= \frac{c_l}{2}(1+c_d \cos(\theta+\phi)),\nonumber
\end{align}
where $\theta$ is the Ramsey phase of the first ensemble, and $\phi$ is the relative phase between the two ensembles --- the quantity we are interested in measuring. In the regime where $\theta$ is well controlled, $p_x$ and $p_y$ can be thought of as the Ramsey signal associated with each sub-ensemble. We operate in the regime where atom-laser coherence is lost, such that $\theta$ is randomized. In this case it is still possible to extract $\phi$ from two ensembles that share a common $\theta$, as we will see below. $ c_l $ is the contribution to the Ramsey contrast from loss (decay of the Ramsey signals towards zero), and $ c_d $ is the contribution to the Ramsey contrast from decoherence or dephasing (decay towards $ c_l/2 $). The overall Ramsey contrast is $ c = c_l c_d $. To infer values for $ c_l $ and $ c_d $ from the measured rate of decay of $c$, we fit the $c$ as a function of time to the product of an exponential and a Gaussian. The Gaussian component is presumed to be exclusively due to dephasing, and thus contributes only to $ c_d $. The exponential component is broken up between the theoretically known ratios between processes that remove atoms from the $\{ \ket{e},~\ket{g} \}$ manifold, namely atom loss and Raman scattering from $ \rm ^3P_0 $ to $ \rm ^3P_2 $, and processes that project onto $ \ket{e} $ or $ \ket{g} $, namely black body or Raman scattering from $ \rm ^3P_0 $ to $ \rm ^3P_1 $, and black body scattering from $ \rm ^3P_0 $ back to $ \rm ^3P_0 $. The former processes contribute to $ c_l $, and the latter to $ c_d $.

We now define the quantities $x$ and $y$, which correspond to the measured excitation fraction in the two ensembles offset to be centered at 0:

\begin{align}
    x &= \frac{c}{2}\cos{\theta}\\
    y &= \frac{c}{2}\cos(\theta+\phi)\,.\nonumber
\end{align}
Note that while these quantities depend only on the product of $ c_l $ and $ c_d $, this is not true of their variance, which can be written as:

\begin{align}
\sigma^2(x) &= \frac{1}{N} p_{x}(1-p_{x})\\
&= -\frac{c_l}{4 N} (1+c_d \cos(\theta))(c_l - 2 + c_d c_l \cos(\theta)),\nonumber
\end{align}
where $N\sim75$ is the number of atoms in each ensemble. An analogous expression holds for $\sigma^2(y)$. Based on this, we can infer a variance in our measurement of $\phi$:

\begin{align}
\sigma^2(\phi) &= \left\lvert\frac{\partial\phi}{\partial x}\right\rvert^2 \sigma^2(x) + \left\lvert\frac{\partial\phi}{\partial y}\right\rvert^2 \sigma^2(y)\\
&= \frac{4}{c^2}\left(\csc^2(\theta)\sigma^2(x) + \csc^2(\theta + \phi)\sigma^2(y)\right).\nonumber
\end{align}
Because we operate with Ramsey dark times far longer than the atom-laser coherence time, we effectively average over all Ramsey phases $\theta$. In this case:

\begin{align}
    \langle{\sigma^2(\phi)}\rangle &= \frac{4}{c^2}\left(\int_0^{2\pi}\frac{d\theta}{2\pi}\frac{1}{\csc^2(\theta) \sigma^2(x)+\csc^2(\theta+\phi) \sigma^2(y)}\right)^{-1},
\end{align}
In the limiting case where $c=1$, QPN results in a variance of $\langle{\sigma^2(\phi)}\rangle = 2/(N c^2)$ as expected. For $c<1$ and as $\phi$ approaches $\pi/2$ however, $\langle{\sigma^2(\phi)}\rangle$ increases by an additional scale factor.

This would suggest that to maximize sensitivity it would be beneficial to operate near $\phi=0$, however, as stated in the main text, the ellipse fitting procedure is biased near this phase. This is due to a combination of two effects: first, these fits cannot distinguish between positive and negative values of $\phi$, and only return $\phi\ge0$. Second, the effective probability distribution the data is sampled from is the convolution of an ellipse and a Gaussian distribution associated with QPN. As a result, for $\phi\simeq0$, data points can cross the semi-major axis due to QPN. In this case, due to the first effect, instead of cancelling with points that cross over from the other side, this results in a weighting away from $\phi=0$. Based on this reasoning, a practical requirement is that $\phi$ is sufficiently large for two peaks to be resolvable along a slice through the semi-minor axis of the ellipse. A similar line of reasoning holds for very long interrogation times, where the contrast of the signal becomes small enough that points can cross the semi-minor axis of the ellipse due to QPN.

\begin{figure}[!htb]
    \includegraphics[width=0.45\linewidth]{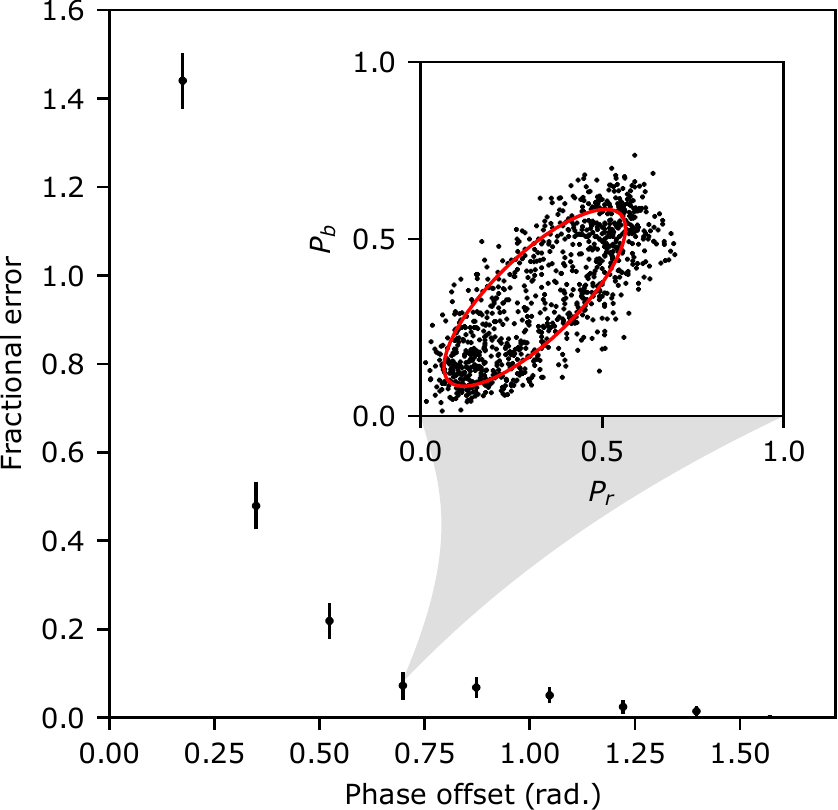}
    \caption{Quantifying bias in ellipse fitting. By simulating data with a known phase offset, we can compute the fractional error in the phase extracted via ellipse fitting. Callout shows an example of a simulated data set with conditions similar to those used for the measurement of relative stability in the main text, where this error is bounded to below 10\%.}
    \label{fig:monte_carlo}
\end{figure}

To quantify the effect of this biasing, we generate artificial data with a known contrast and phase offset via Monte-Carlo simulations that capture the QPN and measured decoherence in our system. By comparing the phase and contrast extracted via ellipse fitting to the known parameters used to generate the data, it is possible to extract a fractional error due to biasing (Fig.~\ref{fig:monte_carlo}). For measuring the decay of Ramsey contrast, we use this error model to correct the data and provide a bias-free estimate of the contrast. For measurements of the phase, we are simply interested in providing bounds on this error to ensure that we operate in an unbiased regime. At $\phi>0.65$ we can bound this error to below 10\%, which is the regime in which we perform the measurement of relative stability in the main text.

To calculate a variance in the extracted value of $\phi$ we turn to the jackknifing procedure described in \cite{marti_imaging_2018}. Specifically, in order to compute an Allan variance it is necessary to have a single-shot estimate of the phase. This is done by checking how much the single-shot measurement of interest, $\phi_i$, sways the estimated phase of the whole distribution:

\begin{equation}
    \phi_i = n \phi-(n-1)\phi_{\neq i}\,,
\end{equation}

\noindent where $n$ denotes the total number of samples, and $\phi_{\neq i}$ is the estimated phase of the distribution with the $i^{\mathrm{th}}$ sample removed. By performing this analysis across all $i$ the Allan variance can be computed in the normal way. The variance in $\phi$ can also be approximated as:

\begin{equation}
    \sigma^2(\phi) = \frac{n-1}{n}\sum_i(\phi - \phi_{\neq i})^2.
\end{equation}

This analysis is used on real data to determine an optimal dark time in the main text, but also on simulated data to calculate a correction factor to the statistical variance of the biased fits compared to QPN. For the conditions of the relative stability measurement in the main text, at $\phi=0$ this results in a correction factor of 0.70(1). For $\phi>0.65$ this analysis yields results bounded to within 5\% of the correct value. This extra source of error is included in the error bars on the relative fractional frequency stability quoted in the main text. We also confirm that including the fringes observed in Fig.~\ref{fig:correlation} of the main text does not alter the results of this analysis --- simulated data with and without these fringes yield results that are within error bars of each other, as is clarified in supplement section \ref{sec:fringes}.

\section{Measuring atomic coherence with the correlation function}\label{sec:correlations}

In this section, we define the correlator used in the main text, and show how it can be used to quantify atomic coherence. We will use the ordered basis $\{ \ket{e}, \ket{g}, \ket{0} \}$ to describe the state of each atom, where $\ket{e}$ and $\ket{g}$ correspond to the $\rm^3P_0$ and $\rm ^1S_0$ states respectively, and $\ket{0}$ corresponds to either a dark state (e.g. $\rm^3P_2$), or the absence of an atom in a given tweezer. Our measurements do not distinguish between $\ket{g}$ and $\ket{0}$, such that we are always measuring the observable:

\begin{align}
\sigma^z = 
\begin{pmatrix}
1&0&0\\
0&-1&0\\
0&0&-1
\end{pmatrix}.
\end{align}
Given these definitions, we define the correlation function between the measured state of two atoms in tweezers $i$ and $j$ as:

\begin{align}\label{eq:corr}
g_{ij}^{(2)} &= \langle(\sigma^z_j - \langle\sigma^z_j\rangle) \otimes (\sigma^z_i - \langle\sigma^z_i\rangle)\rangle \\ 
&= \langle\sigma^z_j \otimes \sigma^z_i\rangle - \langle\sigma^z_j\rangle\langle\sigma^z_i\rangle,\nonumber
\end{align}
where the angled brackets denote expectation values. In practice, these brackets denote an average over runs of the experiment that, after initial loading, included an atom in both sites $i$ and $j$. This ensures that fluctuations in the number or distribution of tweezers that are filled with atoms only affect the number of samples and not the value of the correlator. Note that given our fill fraction of $\sim0.45$, the probability of any given pair being populated on a single shot is $(0.45)^2\simeq0.2$.

In the main text, we are interested in averaging $g_{ij}^{(2)}$ over different ensembles $\{i,j\}$ of interest. Specifically, we define $C(\Delta\pvec{r})$ as the average of $g_{ij}^{(2)}$ over all atom pairs $\{i,j\}$ that are separated by a displacement vector $\Delta\pvec{r}$. Similarly, we define $C_{n\times n}$ as the average of $g_{ij}^{(2)}$ over all pairs of atoms in an $n\times n$ block, also averaged over all such blocks contained in the array for improved statistics. Note that in the limit of an infinitely large array (i.e. ignoring edge effects), $C_{n\times n} = \langle C(\Delta\pvec{r})\rangle_{|\Delta r_x|,\,|\Delta r_y|\le n}$. In the following sections, we will make explicit the connection between the measured quantities defined above, and the off-diagonal coherences in the single atom density matrix.

\droptocpage
\subsubsection*{Ramsey pulse and coherent evolution}

To begin, we consider the evolution of atoms under the Ramsey sequence employed in our experiment. After initial loading, a tweezer $j$ containing an atom has the density matrix $\rho^0_j = \ket{g}\bra{g}$. We perform the first of two SU(2) rotations by applying our clock laser to the atoms. This can be represented by the unitary operator $U_1$:

\begin{align}\label{eq:U1}
U_1 = 
\begin{pmatrix}
\cos(\alpha/2)& -i \sin(\alpha/2)&0\\
-i \sin(\alpha/2)&\cos(\alpha/2)&0\\
0&0&1
\end{pmatrix},
\end{align}
where we have chosen a reference frame in which the rotation occurs about the $x$-axis; and $\alpha$ is the rotation angle, which would be $\pi/2$ in an ideal sequence. This transforms the system into the state $\rho^\text{1}_j = U_1 \rho^0_j U_1^{\dagger}$. The atom is then held in this state for a variable dark time. Due to the energy splitting between $\ket{g}$ and $\ket{e}$, this applies the unitary operator $U^{\text{hold}}_j$:

\begin{align}
U^{\text{hold}}_j = 
\begin{pmatrix}
e^{i\phi_j/2}&0&0\\
0&e^{-i\phi_j/2}&0\\
0&0&1
\end{pmatrix}. \label{eq:Uhold}
\end{align}
This evolution is indexed by the tweezer site $j$ in order to account for spatially inhomogeneous light shifts. Assuming perfect unitary evolution, this leaves that atom in the state $\rho^\text{R}_j$:

\begin{align}
\rho^\text{R}_j= \begin{pmatrix}
 \sin^2(\alpha/2)& -i e^{i\phi_j}\cos(\alpha/2) \sin(\alpha/2)&0\\
i e^{-i\phi_j} \cos(\alpha/2) \sin(\alpha/2)& \cos^2(\alpha/2)&0\\
0&0&0
\end{pmatrix}.\label{eq:rhoR}
\end{align}
However, throughout this sequence the atom can also be projected into other states via the scattering and loss mechanisms discussed in other sections. Since these processes are either incoherent, or drive the atom into the $\ket{0}$ state, we can model their effects by taking the state at the end of this sequence to be a statistical mixture of $\rho^{\text{R}}_j$ and the three basis states. This new state is defined as $\tilde{\rho}_j$:

\begin{align}
\tilde{\rho}_j = p^{(\rm{R})} \rho^\text{R}_j + p^{(g)} \ket{g}\bra{g} + p^{(e)} \ket{e}\bra{e} + p^{(0)} \ket{0}\bra{0},\label{eq:rhoTwid}
\end{align}
where $p^{(\rm{R})},~p^{(e)},~p^{(g)},$ and $p^{(0)}$ are the classical probabilities associated with each of these states, and sum to 1. We label the off diagonal of the single-particle density matrix, $|\rho_{eg}|$, as the single-atom coherence. For perfect $\pi/2$ pulses $\alpha$, $2|\rho_{eg}| = p^{(\rm R)}$ is equal to the Ramsey contrast $c$. In the limit of no dephasing, the decay of the ensemble average of $|\rho_{eg}|$ or $p^{(\rm R)}$ captures the rate at which decoherence occurs due to projection onto any of the states $\ket{0}, \ket{g}$, or $\ket{e}$, from atom loss, spontaneous emission, BBR, or Raman scattering.

\subsubsection*{Measurement protocol}

In order to complete the Ramsey sequence, we apply another SU(2) rotation. However, for long hold times the loss of atom-laser coherence means that this rotation occurs about a random axis on the equator of the Bloch sphere. This rotation, which we call $U_2$, can be written as:

\begin{align}
&U_2 =
\begin{pmatrix}
\cos(\alpha/2)&-ie^{-i\theta}\sin(\alpha/2)&0\\
-ie^{i\theta}\sin(\alpha/2)&\cos(\alpha/2)&0\\
0&0&1
\end{pmatrix}, \label{eq:U2}
\end{align}
where $\theta$ parametrizes the rotation axis $\hat{n} = \cos(\theta) \hat{x} + \sin(\theta) \hat{y}$. This equation assumes that the second rotation sweeps out the same angle $\alpha$ as the first. The final density matrix that we measure, $\rho_j = U_2 \tilde{\rho}_j U_2^{\dagger}$, has five non-zero elements:

\begin{align}
\rho^{00}_j =&~ p^{(\rm{R})} [ 2\sin^2(\alpha/2)\cos^2(\alpha/2)(\cos(\phi_j + \theta) + 1) ]  + p^{(e)}\cos^2(\alpha/2) + p^{(g)}\sin^2(\alpha/2) \\
\nonumber \\
\rho^{01}_j =&~ ie^{-i\theta}[p^{(\rm{R})} \sin^3(\alpha/2)\cos(\alpha/2)(1 + e^{-i(\phi_j + \theta)}) 
- p^{(\rm{R})} \cos^3(\alpha/2)\sin(\alpha/2)(1 + e^{i(\phi_j + \theta)})\nonumber \\
&+ \sin(\alpha/2)\cos(\alpha/2)(p^{(e)} - p^{(g)})] \nonumber \\
\nonumber \\
\rho^{10}_j =&~ -ie^{i\theta}[p^{(\rm{R})} \sin^3(\alpha/2)\cos(\alpha/2)(1 + e^{i(\phi_j + \theta)})
 - p^{(\rm{R})} \cos^3(\alpha/2)\sin(\alpha/2)(1 + e^{-i(\phi_j + \theta)}) \nonumber \\
&+ \sin(\alpha/2)\cos(\alpha/2)(p^{(e)} - p^{(g)})]\nonumber \\
\nonumber \\
\rho^{11}_j =&~ p^{(\rm{R})} [ \sin^4(\alpha/2) + \cos^4(\alpha/2)
- 2\cos^2(\alpha/2)\sin^2(\alpha/2)\cos(\phi_j + \theta) ] \nonumber \\
& + p^{(g)}\cos^2(\alpha/2) + p^{(e)}\sin^2(\alpha/2) \nonumber \\
\nonumber \\
\rho^{22}_j =&~ p^{(0)}.\nonumber
\end{align}
The main quantity we are interested in measuring is the correlator $ g_{ij}^{(2)} $, for which we need the density matrix associated with the full tweezer-array. Since the atoms are not entangled, this is simply $\rho = \bigotimes_{i=0}^{N-1} \rho_i$, where $N$ is the number of tweezers in the array. 

To write down an expression for $g_{ij}^{(2)}$ we first compute $\langle \sigma^z_j\rangle_\theta$, where $\theta$ serves as a reminder that we are really sampling from a compound probability distribution in which $\theta$ is a uniformly distributed random variable. In this case we have:

\begin{align}
\langle \sigma^z_j\rangle_\theta &= \rho^{00}_j - \rho^{11}_j - \rho^{22}_j\\
&=p^{(\rm{R})} [ \sin^2(\alpha) \cos(\phi_j + \theta) - \cos^2(\alpha) ] 
+ (p^{(e)} - p^{(g)} )\cos(\alpha) - p^{(0)}.\nonumber
\end{align}
Marginalizing over $\theta$ we have:

\begin{align}\label{eq:zMean}
\langle \sigma^z_j\rangle & = \frac{1}{2\pi} \int_0^{2\pi} \langle \sigma^z_j\rangle_\theta \text{d}\theta \\
&= (p^{(e)} - p^{(g)} )\cos(\alpha) - p^{(\rm{R})} \cos^2(\alpha)  - p^{(0)}. \nonumber
\end{align}
Because $\rho$ is a tensor product of the single particle density matrices, we also have $\langle \sigma^z_j \otimes \sigma^z_i \rangle_\theta = \langle \sigma^z_j \rangle_\theta \langle \sigma^z_i \rangle_\theta$, giving:

\begin{align}\label{eq:ztenzMean}
\langle \sigma^z_j \otimes \sigma^z_i \rangle = [p^{(\rm{R})}]^2  \sin^4(\alpha) \frac{\cos(\phi_j - \phi_i)}{2} + \langle \sigma^z_j \rangle \langle \sigma^z_i \rangle.
\end{align}
Plugging Eqns.~\ref{eq:zMean} and \ref{eq:ztenzMean} into Eqn.~\ref{eq:corr} yields:

\begin{equation}
g_{ij}^{(2)} = \frac{1}{2}[p^{(\rm{R})}]^2  \sin^4(\alpha) \cos(\phi_j - \phi_i). \label{eq:gij}
\end{equation}

\subsubsection*{Expectation value of measurement protocol} \label{sec:measurementProtocol}

Restricting our attention to the measurement of $C_{n\times n}$ as defined above, we consider the average correlator between all non-empty tweezers within a block $b$:

\begin{align}
C_b \equiv \frac{1}{N(N-1)}  \sum_{ i\neq j}  g_{ij}^{(2)} ,
\end{align} 
where $N$ is the number of atoms in the block. The average over trials should converge to the value $g_{ij}^{(2)}$ calculated in the previous section, giving:

\begin{align}\label{eq:barGn}
C_b = \frac{[p^{(\rm{R})}]^2  \sin^4(\alpha)}{2 N(N-1)} \sum_{ i \neq j} \cos(\phi_j - \phi_i).
\end{align}
Note that $C_{n\times n}$ is simply the average of $C_b$ over all blocks in the array of the appropriate class, and, to the extent that each block behaves identically, will simply equal $C_b$.

In addition to looking at the correlator between all pairs within a block, we also look at the correlator between atoms loaded into a single tweezer site, and all of the other atoms in the array (Fig.~\ref{fig:correlation}b in main text). If the single atom has index $j$, then the outcome of this analysis yields

\begin{equation} \label{eq:mean_atom}
C_{j,A} = \frac{[p^{(\rm{R})}]^2  \sin^4(\alpha)}{2 (N-1)} \sum_{i \in A} \cos(\phi_i - \phi_j),
\end{equation}
where $A$ contains all indices $i\neq j$ in a reference ensemble of interest. We next define the average phase for the atoms in the ensemble $A$ as $\phi_A$, and $\Delta \phi_i \equiv \phi_i - \phi_A$. Given that the dominant contribution to dephasing is from tweezer-induced frequency shifts, which are symmetrically distributed about the anti-diagonal of the array, we can assume that the $\Delta \phi_i$'s are symmetrically distributed about zero. In this case:

\begin{equation} \label{eq:mean_atom2}
C_{j,A} = \frac{[p^{(\rm{R})}]^2  \sin^4(\alpha)}{2 (N-1)} \cos(\phi_A - \phi_j) \sum_{i \in A} \cos(\Delta \phi_i).
\end{equation}
For large atom number where $A$ is indistinguishable from the full array, $ p^{(\rm{R})} \sin^2(\alpha) \sum_{i \in A} \cos(\Delta \phi_i)$ is equal to the Ramsey contrast $c$, including the effects of imperfect $\pi/2$ pulses and dephasing. $c$ can be independently characterized using the ellipse fitting methods discussed in supplemental section \ref{sec:ellipse}, and divided out of Eqn. \ref{eq:mean_atom2} (as is done in Fig.~\ref{fig:correlation}b of the main text). For atoms near the center of the array where $\phi_j\simeq \phi_A$, this yields a direct measurement of $p^{(\rm{R})} \sin^2(\alpha)$, which is insensitive to dephasing and proportional to the single-atom coherence $|\rho_{eg}|$. Note that if $\phi_j\neq \phi_A$, this measurement decays strictly faster than $|\rho_{eg}|$. In practice, we average $ C_{j,A} $ over a $ 4\times4 $ region at the center of the array, yielding $ C_A $. To extract the single-atom coherence from $ C_A $, we fit it with a fit function that includes a fixed contribution from the decay of $ c $ extracted from ellipse fitting, and a free exponential component. The $ 1/e $ lifetime of this free component is the inferred single-atom coherence time, and has a value of 48(8)~s.

\subsubsection*{Comparison with the average density matrix}

It is possible to show that a measurement of $\sqrt{C_{n\times n}}$ constitutes a lower bound on the atomic coherence. To begin, consider the average density matrix $\bar{\rho}$ of all atoms in a given block:

\begin{align}
\bar{\rho} \equiv \frac{1}{N} \sum_{j=0}^{N-1} \tilde{\rho}_j  &=
\begin{pmatrix}
p^{(e)}&0&0\\
0&p^{(g)}&0\\
0&0&p^{(0)} 
\end{pmatrix} +
\frac{p^{(\rm{R})} \sin(\alpha)}{2}
\begin{pmatrix}
\tan^2(\alpha/2)& \frac{1}{N} \sum_{j=0}^{N-1}  -i e^{i\phi_j} &0\\
 \frac{1}{N} \sum_{j=0}^{N-1} i e^{-i\phi_j} &\cot^2(\alpha/2)&0\\
0&0&0
\end{pmatrix}.
\end{align}

We define the atomic coherence as the magnitude of the off-diagonal element, $|\bar{\rho}_{eg}|$, where:

\begin{align}
|\bar{\rho}_{eg}|^2 &=\frac{( p^{(\rm{R})} \sin(\alpha))^2}{4}  \frac{1}{N} \sum_{j=0}^{N-1}  e^{-i\phi_j} \frac{1}{N} \sum_{j=0}^{N-1}  e^{i\phi_j} \\
&= \Bigl{(} \frac{p^{(\rm{R})} \sin(\alpha) }{2N} \Bigr{)}^2 \ \Bigl{[} \sum_{j,j'} \cos(\phi_j - \phi_{j'}) \Bigr{]} \nonumber.
\end{align}
Writing this in terms of $C_b$ using Eqn.~\ref{eq:barGn} yields:

\begin{align}
|\bar{\rho}_{eg}|^2 = \frac{1}{N} \Bigl{(} \frac{p^{(\rm{R})} \sin(\alpha) }{2} \Bigr{)}^2 +  \frac{1}{\sin^2(\alpha)} \frac{C_b}{2},
\end{align}
which then gives:

\begin{align}\label{eq:bounds}
C_b \leq 2 \sin^2(\alpha)  |\bar{\rho}_{eg}|^2 \leq  2 |\bar{\rho}_{eg}|^2.
\end{align}
The factor of 2 in this inequality is due to the differing normalizations of $|\bar{\rho}_{eg}|$ and $C_{n\times n}$. Specifically, $|\bar{\rho}_{eg}|^2$ ranges from 0 to $1/4$, whereas after marginalizing over the laser phase, $C_{n\times n}$ varies between $-1/2$ and $1/2$ (where we are only concerned with the limit in which phase shifts are small, and $C_{n\times n}>0$).

Eqn.~\ref{eq:bounds} implies that a measurement of $\sqrt{C_b/2}$ is a lower bound on the atomic coherence, since $\sqrt{C_b/2} \leq  |\bar{\rho}_{eg}|$. It is natural to average $\sqrt{C_b/2}$ over all appropriate $b$ in the array, using $\langle \sqrt{C_b/2} \rangle$ to characterize the array-averaged atomic coherence. However, to avoid negative values in the square root due to statistical fluctuations we instead average over $b$ before performing the square root, yielding $\sqrt{C_{n\times n}/2}$ where $C_{n\times n}=\langle C_b \rangle$. Note that by the triangle inequality $\sqrt{C_{n\times n}/2} \le \langle \sqrt{C_b/2} \rangle$, and still constitutes a valid lower bound. Assuming that $\alpha$ does not vary with dark time, the lifetime of $\sqrt{C_{n\times n}}$ similarly provides a lower bound on the atomic coherence time.

\incltocpage
\section{Fringes and spatially varying frequency shifts}\label{sec:fringes}

\begin{figure}[!htb]
    \centering
    \includegraphics[width=0.55\textwidth]{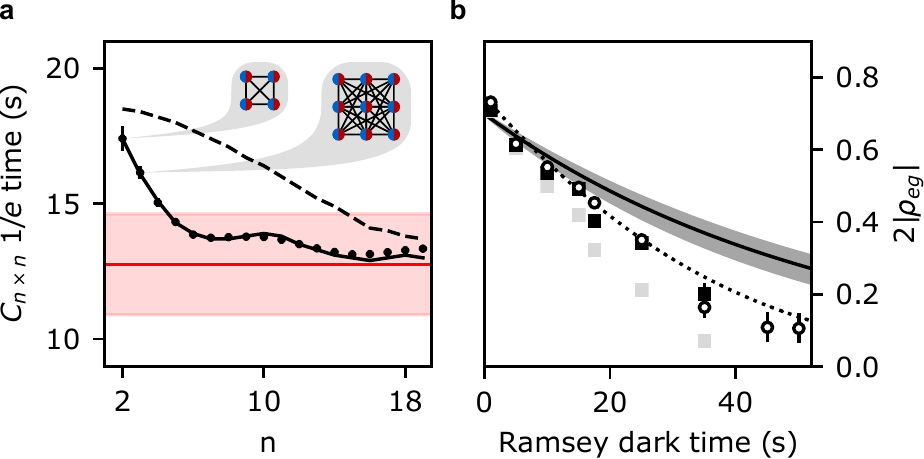}
    \caption{Decay of correlations as a function of sub-ensemble size. a) Extending the analysis of correlations in $2\times2$ blocks of atoms in the main text to $n\times n$ blocks of varying size (see callouts) reveals that the inferred atomic coherence decays faster with increasing block size than the prediction based exclusively on theoretical tweezer frequency shifts and a fitted exponential decay constant (dashed line). An extended model that includes the fringes that appear in the correlator (solid line) is in good agreement with the data. The inferred lifetime based on the ellipse fitting procedure in Fig.~\ref{fig:atomatom} of the main text is shown in red (shaded region denotes $1\sigma$ confidence interval) and is in good agreement with this analysis. Note that for $n>16$, the blocks are rectangular with dimensions $n\times 16$. b) The bound on atomic coherence based on a measurement of $C_{n\times n}$ becomes looser with larger $n$ due to increased sensitivity to dephasing. For example, the atomic coherence inferred from averaging over all correlators, or equivalently $C_{20\times16}$ (grey squares), decays much more quickly than for $C_{2\times 2}$ (open circles, reproduced from Fig.~\ref{fig:correlation}c of the main text). However, based on our knowledge of this dephasing we can remove the contribution from dephasing from the atomic coherence inferred from $C_{20\times16}$, which yields results (black squares) that are in good agreement with the bound set by $C_{2\times 2}$.
    }
    \label{sfig:blockSize}
\end{figure}

We observe fringes in $C(\Delta\pvec{r})$ that grow linearly in amplitude as a function of dark time. To study the effect of these fringes on our measurements we compute the rate at which correlations decay in $n\times n$ sub-ensembles of varying size (see Fig.~\ref{sfig:blockSize}a). For intermediate block sizes, the presence of the fringes results in substantially more rapid decay of the correlator. However, for block sizes that are small compared to the fringes, as is the case for the $2\times 2$ blocks used in the main text, and for large blocks that sample the entire array, the presence of these fringes only marginally increases the decay rate. As a result these fringes do not have a substantial impact on either our ability to extract a single-atom coherence, or on the stability achievable in a synchronous clock comparison employing the entire array. Critically, the excess decay when averaging over the whole array compared to the single-atom coherence is fully explained by dephasing. This is further evidenced by the fact that the $20\times 16$ correlator is in good agreement with the $2\times 2$ correlator when it is corrected for the expected reduction in correlations due to dephasing at each time (Fig.~\ref{sfig:blockSize}b).

The relatively fine spatial structure of these fringes suggests that they are an optical effect, and the fact that they grow with dark time suggest that they are due to light shifts caused by the 813~nm tweezers, since all other beams are off during this period. This is further evidenced by the fact that the fringes are oriented along the more tightly spaced axis of the array, which leads to greater overlap between the tweezers and difficulties reading out, and thus balancing, their depths. Given the known $\sim10$\% inhomogeneities in trap depth in our system, a 200~MHz detuning of the tweezers from the magic condition can explain this effect. Moreover, the magnitude of this effect fluctuates day to day, which is consistent with the fact that the trap laser frequency is known to drift on the 50~MHz scale.

\section{Master equation modeling of atom-atom correlations}

In this section we introduce a full master equation-based model of our system, which we compare to our measurements of $ C_{2\times 2} $. Here, besides the $\ket{e}, \ket{g},$ and $\ket{0}$ states introduced in section \ref{sec:correlations}, we also include four motional states of the atom to account for finite atomic temperature. As a result, the state of each atom is represented by a $12\times12$ density matrix $\rho$. To compare to our measurements of $ g^{(2)} $, we simulate two such atoms. In this case, because there is no mechanism for the development of quantum correlations, each atom can be simulated independently.

The atoms begin in a statistical mixture of $\ket{g}$ and $\ket{0}$ due to a total of 9\% loss during the imaging and hand-off from the auxiliary to the science potential. To simulate our Ramsey measurement, each atom is rotated in the $\{\ket{e}, \ket{g}\}$ subspace by an angle $\alpha$, which is $\pi/2$ for the average thermal ensemble. Specifically, each atom samples a thermal Boltzmann distribution defined by an average motional excitation number of $\overline{n}=0.14$, and the rotation angle is modified according to each atom's sampled motional excitation state $ n $ (see supplement \ref{sec:finiteTemp}). The atoms then undergo free evolution, accruing a phase $\theta$ relative to the laser. They further experience incoherent decay, and coherent dephasing which yields a relative phase $\phi$. For readout, the result is rotated as at the beginning of the sequence, although $ n $ and thus $\alpha$ can be modified by heating due to Rayleigh scattering. With this final state in hand, we can compute expectation values of the relevant operators to obtain our results. 

The evolution of the density matrix $\rho$ is given by a Lindblad master equation
\begin{align}
    \label{eqn:mastereqn}
    \dot{\rho} &= -i[H,\rho] + \sum_i \Gamma_i\mathcal{L}(\rho,\mathcal{P}_i)\\
    \mathcal{L}(\rho,\mathcal{P}) &= \mathcal{P}\rho\mathcal{P}^\dagger - \frac{1}{2}\left(\rho\mathcal{P}^\dagger\mathcal{P} + \mathcal{P}^\dagger\mathcal{P}\rho\right),
\end{align}
where $H$ is the Hamiltonian, $\Gamma_i$ is the rate of the $i^{th}$ incoherent process, and $\mathcal{P}_i$ is the corresponding jump operator. To write down expressions for $ H $ and $\mathcal{P}_i$, we define the single-atom Pauli matrices, which operate on the ground-excited subspace of the internal degrees of freedom:
\begin{equation}
\sigma_x^0 = \left(\begin{tabular}{ccc}
0 & 1 & 0 \\ 
1 & 0 & 0 \\
0 & 0 & 1 \\
\end{tabular}    \right)\,\otimes \, \mathbb{1}_4 \,\, , \,\, 
\sigma_y^0 = \left(\begin{tabular}{ccc}
0 & $-i$ & 0 \\ 
$i$ & 0 & 0 \\
0 & 0 & 1 \\
\end{tabular}    \right)\,\otimes \, \mathbb{1}_4\,\, , \,\,
\sigma_z^0 = \left(\begin{tabular}{ccc}
1 & 0 & 0 \\ 
0 & -1 & 0 \\
0 & 0 & 1 \\
\end{tabular}    \right) \,\otimes \, \mathbb{1}_4\,\, 
\end{equation}
where $\mathbb{1}_4$ is the $4\times4$ identity matrix on the motional degrees of freedom. 

To compare to our measurements of $ C_{2\times 2} $, we simulate pairs of atoms with a differential clock frequency shift that is equal to the mean tweezer-induced frequency difference between atom pairs in a $2\times2$ block. In this case the two atoms evolve under the Hamiltonians $H = \pm\delta_1\sigma_z^0/2$, where $\delta_1 = 2\pi\times 2.4$~mHz (see supplement~\ref{sec:lightshiftDerivation}). Note that $ \delta_1 $ includes the effects of the fringes observed in the correlator (see Fig.~\ref{fig:correlation}a in the main text, and Fig.~\ref{sfig:blockSize}).

The values of $\Gamma_i$ are inferred from the measured decay rates presented in section \ref{sec:clockLifetime}, and the theoretically known relative rates and branching ratios between the various relevant processes. Specifically, given the measured excited-state decay rate $\Gamma_e$, loss rate of excited-state atoms $\Gamma^t_{e}$, and independently characterized black body decay rate $\Gamma^{\mathrm{BBR}}_1$~\cite{dorscher_lattice-induced_2018}, the total Raman scattering rate due to the science potential is $\Gamma^{\rm R}_{12} = \Gamma_{e} - \Gamma^t_{e}-\Gamma^{\mathrm{BBR}}_1$. Given the relative rates of the Raman (and Rayleigh) scattering processes of $R^S_J = 0.416$, $0.372$, and $0.212$ for $J = $ 0, 1, and 2 (where these rates are normalized to the total scattering rate), we can infer the individual process rates $\Gamma^{\rm R}_{J} = \Gamma^{\rm R}_{12} R^S_J/(R^S_1+R^S_2)$.

Based on the above definitions, we incorporate decay from the excited to ground state by setting $\Gamma = \Gamma^{\rm R}_{1} + \Gamma^{\mathrm{BBR}}_1 = 8.7 \cdot 10^{-3}$ s$^{-1}$ and the corresponding jump operator to be $\sigma^{-} = \frac{1}{2}\left(\sigma_x^0 - i\sigma_y^0\right)$. We incorporate atomic loss from the ground and excited states by setting $\Gamma^L_g = \Gamma^t_{g} = 9.9\cdot10^{-3}$ s$^{-1}$ and $\Gamma^L_e = \Gamma^t_{e} + \Gamma^{\rm R}_{2} = 14.6\cdot10^{-3}$ s$^{-1}$, with the corresponding jump operators $\sigma_{g0} = \ket{0}\bra{g} \otimes \mathbb{1}_4$ and $\sigma_{e0} = \ket{0}\bra{e} \otimes \mathbb{1}_4$. We incorporate blackbody Rayleigh scattering off the excited state driving the $\rm^3P_0 \rightarrow \,^3D_1 \rightarrow \,^3P_0$ process by using $\Gamma^{\rm BBR}_0 = \Gamma^{\mathrm{BBR}}_1 R^D_0/R^D_1 = 3.45\cdot 10^{-3}$ s$^{-1}$ and the corresponding jump operator to be $\sigma_{ee} = \ket{e}\bra{e} \otimes \mathbb{1}_4$. Finally, we incorporate heating due to Rayleigh scattering of the 813~nm tweezer light by setting $\Gamma^H = 2\Gamma^{\rm R}_{0}/3 = 4.8\cdot 10^{-3}$ s$^{-1}$ and the corresponding jump operator to be $\sigma_H = \mathbb{1}_3 \otimes \exp\left(i\eta(a+a^\dagger)\right)$, where $a$ ($a^\dagger)$ is the motional lowering (raising) operator. The factor of $2/3$ accounts for two photons scattering per event and all axes thermalizing. Note that Rayleigh scattering of the tweezer light does not directly introduce dephasing since the scattered photon carries no information about which internal state the atom was in~\cite{dorscher_lattice-induced_2018}.

\begin{figure}[!tb]
    \centering
    \includegraphics[width=0.9\textwidth]{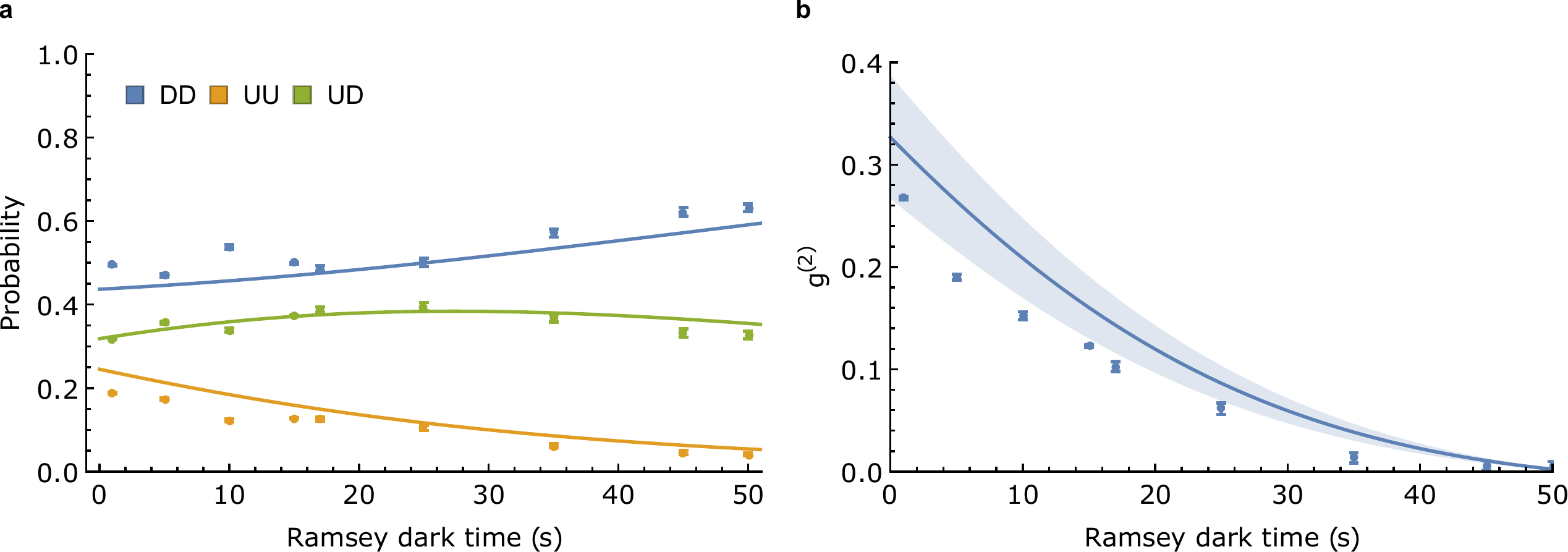}
    \caption{Results of the master equation model. (a) The observed population ratios of UU, UD, and DD are in reasonable agreement with the prediction of our model with no free parameters. (b) We compare the observed $ 2\times2 $ block correlator, $ C_{2\times2} $ (blue points, see also Fig.~\ref{fig:correlation} of the main text) to the prediction of our model with no free parameters. This model predicts a correlation 1/$e$ time of 19.7(1.2) seconds, somewhat longer than the observed 1/$e$ time of 17.4(9) seconds (this value includes a Gaussian and an exponential component, and is quoted in the main text as 33(2)~s, associated with the decay of $ \sqrt{C_{2\times2}} $). The errorband propagates uncertainty in the $\rm^3P_0$ state lifetime and initial temperature; the uncertainty in the ground state lifetime is negligible. If we do not assume that all three axes thermalize after Rayleigh heating, such that $\Gamma_H = \Gamma_{\rm R}^0/3$, the model's 1/$ e $ time increases by 5\% while the initial value remains approximately constant.}
    \label{fig:pairpops}
\end{figure}

We numerically solve this master equation for each atom, and run the simulation for ten different relative phase shifts $\theta$ uniformly distributed between $0$ and $2\pi$ to account for experimental averaging over the laser phase; we find that increasing the number of laser phases simulated does not affect our results.

As a function of time, we compute the average $g^{(2)}$ correlator as well as the expectation values of other observables corresponding to experimentally accessible two atom populations. Each run of our experiment returns a pair of images, which may be binarized into a pair of $20\times16$ arrays of ones and zeros corresponding to the presence or lack of an atom respectively. The first image corresponds to all atoms present in the array before the first $\pi/2$ pulse, while the second image shows atoms in the excited clock state after the second $\pi/2$ pulse. We do not distinguish between an atom in the ground state and the atom being lost. Therefore, we construct the observables
\begin{subequations}
\begin{align}
    UU &= P_e \otimes P_e  \\
    UD &= P_e \otimes (P_g + P_0) + (P_g + P_0) \otimes P_e \\
    DD &= (P_g + P_0) \otimes (P_g + P_0),
\end{align}
\end{subequations} 
where the projectors are $P_e = \ket{e}\bra{e}$, $P_g = \ket{g}\bra{g}$, and $P_0 = \ket{0}\bra{0}$. As a test of our master equation model, in Fig.~\ref{fig:pairpops}a we compare the expectation values of these three observables between the model and the experimental data, finding close agreement with no free parameters. However, the model slightly overestimates the decay time and initial value of the correlator $ g^{(2)} $ (Fig.~\ref{fig:pairpops}b). This reflects that the correlator is much more sensitive to model parameters than any of the pair population results because it is equal to a difference between quantities of similar magnitude, $g^{(2)} = \langle UU\rangle +\langle DD\rangle- \langle UD\rangle-\langle\sigma_z\rangle_i\langle\sigma_z\rangle_j$ (where $ i $ and $ j $ label the two atoms under measurement, and $ \sigma_z $ is as defined in section \ref{sec:correlations}). Nevertheless, this discrepancy could indicate slightly higher decay rates than independently calibrated, higher atomic temperature of around $\bar{n}=0.25$, increased imaging or handoff loss, or some new and uncharacterized decoherence mechanism.

\section{Finite temperature effects}\label{sec:finiteTemp}

\begin{figure}[!htb]
    \includegraphics[width=0.5\linewidth]{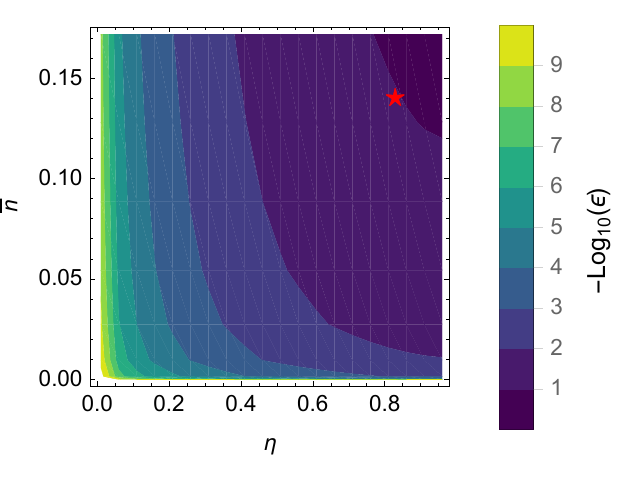}
    \caption{Bounds on clock $\pi$-pulse fidelity due to finite temperature. Predicted $\pi$-pulse error $\epsilon$ as a function of Lamb-Dicke parameter $\eta$ and average phonon occupation $\Bar{n}$ for Rabi and trap frequencies used in this work. Red star denotes conditions used for Ramsey evolution and synchronous clock comparison in main text.}
    \label{fig:fidelity}
\end{figure}

To calculate the expected clock $\pi$-pulse fidelity achievable given our known cooling performance and confinement, we solve for the time evolution of a harmonically trapped atom under a resonant optical drive. Specifically, we consider the interaction Hamiltonian:

\begin{equation}
    H_{int} = \frac{\Omega}{2}(e^{i \eta (a^\dag+a)}S_+ + H.C.)\,,
\end{equation}

\noindent where $\Omega$ is the clock Rabi frequency, $\eta$ is the Lamb-Dicke parameter, $a^\dag$ ($a$) is the motional raising (lowering) operator, $S_+$ is the spin raising operator, and $H.C.$ denotes the Hermitian conjugate of the preceding terms. An atom in a given motional eigenstate $ n $ evolves under this Hamiltonian with an effective Rabi frequency:

\begin{equation}
	\Omega{(n)} = \Omega e^{-\eta^2/2} L_n(\eta^2),
\end{equation}

\noindent where $ L_n $ is the Laguerre polynomial of degree $ n $. We can average this evolution over a thermal ensemble representative of the measured atom temperatures in our experiment to model the time evolution and thus optimal $ \pi $-pulse fidelities achievable in our apparatus.

The 15$\mathrm{E_R}$ deep traps used for Ramsey spectroscopy in the main text result in a trap frequency of $f_\mathrm{trap} = 6.8$~kHz and $\eta = 0.83$. Based on sideband thermometry (see main text) we infer an average phonon occupation of $\Bar{n} \simeq 0.14$ along the direction of the clock probe, corresponding to a temperature of $T = 156$~nK. At this temperature,  we find that the maximum $\pi$-pulse fidelity is limited to 0.90, in good agreement with the measured value of 0.82(2) when further including the $ \sim9\% $ additional losses due to imaging and imperfect calibration of the handoff procedure for that data.

While this limited contrast does not heavily affect clock operation, for applications that require high fidelity transfer to the clock state either the cooling performance or optical confinement must be improved. Fig.~\ref{fig:fidelity} shows the achievable $\pi$-pulse fidelity for different combinations of cooling performance and confinement. One route to improving $\pi$-pulse fidelity is by increasing the confinement of the atoms via an additional optical lattice at 813~nm, which can provide tightly confining potentials with reduced requirements on optical power. A readily achievable trap frequency of 100~kHz in such a lattice would correspond to $\eta\simeq 0.2$, which, assuming the cooling performance achieved in this work of $\Bar{n}\simeq 0.14$ would result in a fidelity of 99.94\%. Note that if desired, the atoms could be transferred into a shallow tweezer potential after such a pulse is completed to make use of the long lifetimes demonstrated in this work.

\section{Limits on scaling}

Our current apparatus is limited by thermal lensing in the optical rail used to project the deep 515~nm tweezers, which limits the usable power in the rail to $\sim1.5$~W. Given the optical power available at 515 and 813~nm, and the RF bandwidth of the AODs used in the tweezer rail, through more careful material selection and optical design such a system could readily be scaled to more than 1000 traps.

Moreover, the approach to scaling laid out in this work is generally applicable to experiments that want to reduce the effects of scattering by using a far-detuned science potential, while using a less far-detuned potential for fast, power hungry stages of the experiment which can alleviate constraints on atom number and/or laser power.

\end{document}